\documentclass[trackchanges]{aastex701}

\usepackage{bm}
\usepackage{comment}

\usepackage{pgfplots}
\usepackage{tikz-3dplot}
 
\tdplotsetmaincoords{60}{115}
\pgfplotsset{compat=newest}

\usepackage{tikz}
\usetikzlibrary{arrows.meta,calc,angles,quotes}

\begin{document}

\title{X-ray polarization in magnetized neutron stars}

\author[orcid=0000-0002-3033-5843]{Tanuman Ghosh}
\affiliation{Inter-University Centre for Astronomy and Astrophysics, Post Bag 4, Ganeshkhind, Pune 411007, India}
\email[show]{tanuman.ghosh@iucaa.in}

\author{Shiv Sethi}
\affiliation{Astronomy and Astrophysics, Raman Research Institute, C. V. Raman Avenue, Sadashivanagar, Bangalore 560080, India}
\email[show]{sethi@rri.res.in}

\begin{abstract}

X-ray polarimetry has opened a new window into understanding the physics around magnetized compact objects. IXPE detection of linear polarization from such systems has prompted a new spurt of theoretical modeling. Our study is based on the dominant paradigm that the observed polarization arises from the scattering of photons around highly magnetized systems. Our main focus is the dependence of the polarization of the scattered light on properties of  the incoming light, i.e., geometry and the polarization state, and the determination of the spectral shape of the polarized light for a wide range of magnetic field strengths. We also analyze the impact of vacuum birefringence on photon polarization. We show that, generically, we expect a higher linear degree of polarization from magnetars as compared to normal pulsars, which is in agreement with IXPE observations. Under some conditions, our study helps to understand the observed degree of polarization from normal pulsars and low-magnetized neutron stars and their spectral dependence. However, we cannot conclusively explain the spectral shape of the observed polarization for magnetars using only a single component emission from scattering in a strong magnetic field. This probably points to the system being more complex, e.g., multi-component, than our study allows for.  Upcoming X-ray polarimeters with broader energy coverage could probe some of our other predictions, e.g., the spectral shape of the polarized light close to the resonance frequency.

\end{abstract}



\section{Introduction} \label{sec: introduction}
Broadband X-ray spectroscopy has been one of the primary mechanisms to understand the physical processes in neutron star (NS) systems. The advent of X-ray polarimetry, particularly with IXPE \citep{Weisskopf2022} in the past decade, has spurred extensive studies on theoretical modeling to understand the observed features of X-ray polarization (e.g., \citealt{Doroshenko2022, Taverna2022, Zane2023, Forsblom2023, Tsygankov2023, Farinelli2023, Ursini2023, Capitanio2023}).

In terms of  the surface magnetic field strength, we can broadly classify these sources into  three categories: low magnetic field NSs with surface fields in the range  $10^7\hbox{--}10^9$~G, which are found in low-mass X-ray binaries and rarely show pulsations in X-rays, highly-magnetized NSs with surface field around $10^{12}\hbox{--}10^{13}$~G,  consistent with normal X-ray pulsars in high-mass X-ray binaries, and very strongly magnetized NSs with surface field $\gtrsim 10^{14}$~G, the magnetars. 

On the theoretical front, many emission mechanisms and geometries have been studied to interpret radiation from these sources. 
Two dominant emission geometries for normal pulsars, where an accretion column is formed, are the pencil and the fan beam. In the former the emission occurs close to the magnetic field
lines while it is normal to the magnetic field orientation in the latter case (see e.g., \citealt{Becker2012} and references therein).
Such geometries have direct bearing on the observed polarization state of the photons (e.g., \citealt{Caiazzo2021, Gnedin1978}).
The  theoretical models  of  such highly magnetized accreting neutron star systems, e.g., high-mass X-ray binaries and normal pulsars, have generically predicted high degrees of linear polarization (e.g., \citealt{Caiazzo2021, Meszaros1988}). Nevertheless,
there are other theoretical complications in modeling such systems. In the presence of a strong magnetic field, the ambient vacuum becomes birefringent, i.e, photons of different polarization states propagate through the medium at different speeds. In a neutron star atmosphere, where particle number density can be very high, the plasma also contributes to the dielectric tensor along with vacuum birefringence.  Several works have theoretically studied the effects of vacuum birefringence and plasma resonance on the polarization of outgoing photons in the presence of a strong magnetic field (e.g., \citealt{1983JOSA...73.1719K, 1997JPhA...30.6485H, 2003ApJ...588..962L, 2003PhRvL..91g1101L, Gnedin1978}).

In the most recent past, IXPE observations  in the energy range 2--8~keV have allowed a comparison of theoretical 
predictions with X-ray polarization data. 
For normal pulsars, most IXPE observations have detected  significantly lower polarization degree  than the expectation of theoretical models \citep{Poutanen2024}.  For magnetars, on the other hand, high polarization degree has been detected (e.g., \citealt{Taverna2024,Taverna_2026}). This provides additional motivation to investigate the emission processes in these systems further.

The polarization of photons from neutrons star systems can primarily arise from two mechanisms: synchrotron radiation or scattering of photons. 
In this paper, we consider the scattering of photons as being responsible for the polarized emission; the dominant physical process in this setting is Thompson scattering in the presence of strong magnetic field (e.g., \cite{1986Ap&SS.121..333C, Caiazzo2021} and references therein). In the vicinity of the neutron star surface, 
the polarization of scattered photons is sensitive to the geometry of scattering, which itself is largely determined by
the distribution of electrons.  A significant complication in relating the polarization of scattered photons with the observed 
photons arises from vacuum and plasma polarization.  These effects can mix different Stokes' parameters and 
might act to depolarize the scattered radiation under some conditions (e.g., \citealt{Caiazzo2021, Gnedin1978, Meszaros1988}).  

In this work, we model the scattering of photons of arbitrary initial polarization in a strong magnetic field, with particular focus on the geometry of the electron distribution. In addition, we also study the possible impact of vacuum polarization on scattered photons. We consider magnetic fields of strengths $10^8\hbox{--}5\times 10^{14} \, \rm G$ in our analysis, which allows coverage from low-magnetic field NSs to extreme objects such as magnetars. 

In the next section, we discuss the physics of the scattering process with a focus on the polarization state of the scattered
photon and its dependence on the geometry of the electron distribution. In \S~\ref{sec:vacuum_birefringence}, we describe in detail
the possible impact of vacuum and plasma birefringence. In \S~\ref{sec:discussion},  we summarize our main findings and discuss possible implications of our work to the current and future data and theoretical modeling of NS systems.

\section{Scattering of photons and neutron star geometry} \label{sec:scattering}

\subsection{Geometric construct} \label{sec:geocon}
\begin{figure}[t]
\centering
\includegraphics[width=0.5\linewidth, keepaspectratio]{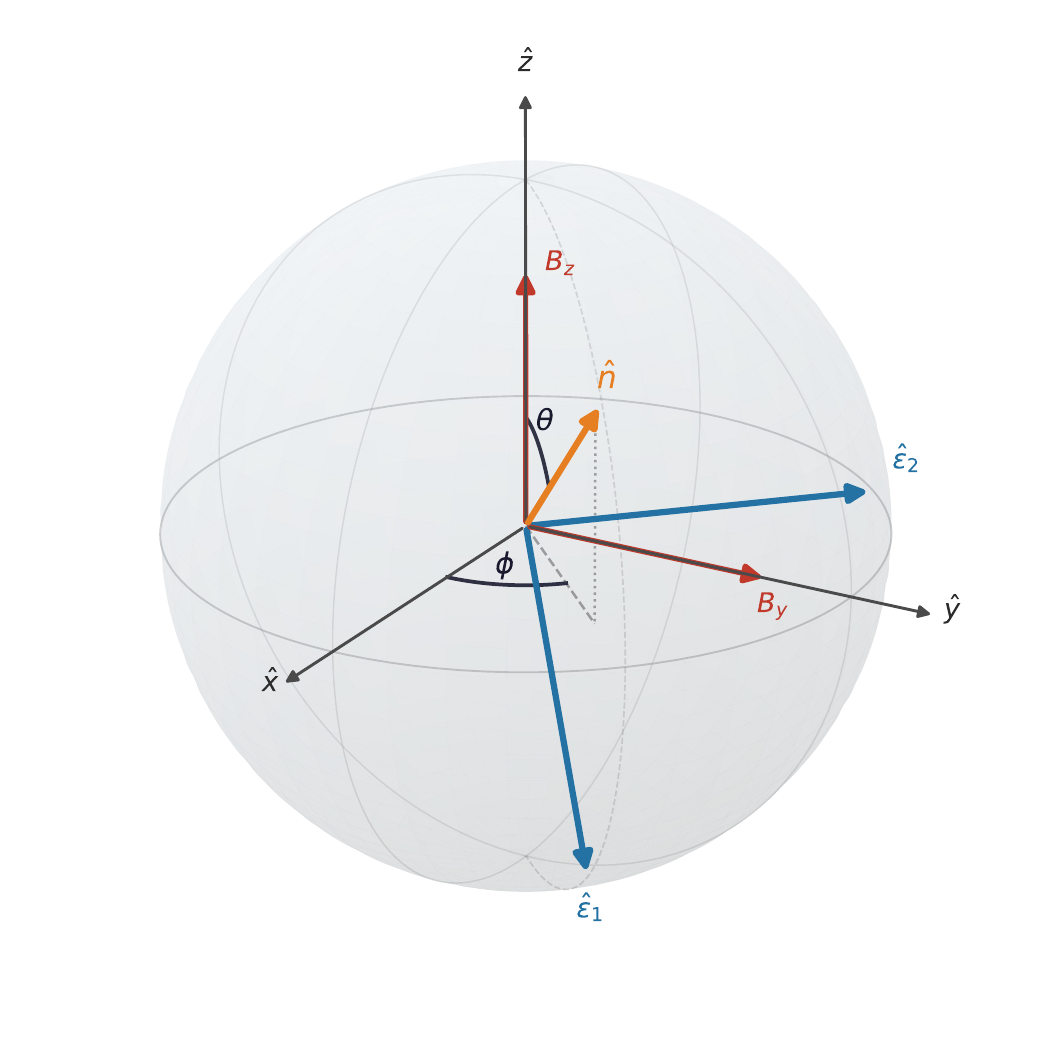}
\caption{Schematic representation of the geometry of the problem. The incoming photon direction $\hat{n}$, polarization vectors $\hat{\epsilon}_1$ and $\hat{\epsilon}_2$, and magnetic field components $B_y$ and $B_z$ are shown. The angles $\theta$ and $\phi$ are marked as the polar and azimuthal angles, respectively.}
\label{fig:vector_geometry}
\end{figure}

We consider the geometry of scattering shown in Figure~\ref{fig:vector_geometry}.
Without loss of generality, the incoming photon of frequency $\omega$ arrives from a direction $\hat n$ in a spherical coordinate system, while the scattered photon is in the $z$-direction. The magnetic field is assumed to lie in the $y\hbox{-}z$ plane. This allows us to express the magnetic field and electric field on incoming photons as: 
\begin{eqnarray}
{\bf B} & = &  B_z \hat z+ B_y \hat y \\
{\bf E} & =  & \left (E_{01} \hat \epsilon_1 + E_{02} \hat \epsilon_2 \right ) \exp(i\omega t)
\end{eqnarray}
with 
\begin{eqnarray}
\hat \epsilon_1 & =  & \cos\theta \cos\phi \hat x + \cos\theta \sin\phi \hat y - \sin\theta \hat z \\
\hat \epsilon_2 & = & -\sin\phi \hat x + \cos\phi \hat y  \\
\hat n  & = & \sin\theta \cos\phi  \hat x + \sin\theta \sin\phi  \hat y +  \cos\theta \hat z
\end{eqnarray}
The direction of the incoming photon $\hat n$, $\hat \epsilon_1$, and $\hat \epsilon_2$ form a set of three orthonormal vectors (Figure \ref{fig:vector_geometry}).   In general, both $E_{01}$ and $E_{02}$ are complex numbers. For fully
polarized light,  both $E_{01}$ and $E_{02}$ are constants. However, for representing partially polarized light, we could treat them
as time-dependent. For instance, for representing unpolarized light, $\langle E_{01} E_{02} \rangle = 0$, where the average is 
over time.

\subsection{Solving equation of motion}
In the presence of a magnetic field, the classical photon scattering off a non-relativistic electron (we assume the electron to be non-relativistic in our treatment) has a resonance when the angular frequency of the photon $\omega$ is close to the gyration frequency $\omega_B = eB/mc$ (e.g., \citealt{1971PhRvD...3.2303C}). To mitigate this singularity, one needs to invoke the semi-classical approach  that allows for the quantum correction to the cross section close to the resonance (e.g., \citealt{1979rpa..book.....R}). The decay constant, $\Gamma$, for electron motion in a magnetic field corresponds to the decay of Landau levels (for details of the quantum mechanical treatment of photon absorption and compton scattering in a strong magnetic field, see e.g.  \citealt{1978PhRvD..18.1053D})\footnote{The dominant decay arises from the decay of lower Landau levels and depending on the magnetic field strength could correspond to same spin or the spin-flip transition.  The former dominates the case for $B \ll B_{\rm crit}$ and gives   $\Gamma = 4\alpha/3 (B/B_{\rm crit})^2 (m_e c^2/\hbar)$; here $B_{\rm crit} = m_e^2 c^3/(e\hbar) = 4.4 \times 10^{13} \, \rm G$. The general case is more complicated and discussed in e.g., \citet{1978PhRvD..18.1053D, Harding1991}. For our purposes,   $\Gamma \ll \omega, \omega_b$. We also show later that $\Gamma$ doesn't play any role in the final expression of polarization of scattered photons.}

The equation of motion of the particle, including the decay term, is:
\begin{equation}
m \mathbf{\dot v} = {e\over c} \mathbf{v}\times\mathbf{B} + e \mathbf{E} - m \Gamma \mathbf{v}
\label{eq:eqofmo_decay}
\end{equation}

We seek the following solution to Eq.~(\ref{eq:eqofmo_decay}):
\begin{equation}
{\bf v} = \left (v_x \hat x + v_y \hat y+ v_z \hat z \right ) \exp(i\omega t)
\end{equation}

This gives us: 
\begin{eqnarray}
v_{x\epsilon_1}  & = &   (1- \omega_b^2/\omega^2-2i\Gamma\omega_b^2/\omega^3)^{-1} \left(  eE_{01} \left[-{\omega_{by}\over m\omega^2} \sin\theta + {\cos\theta \cos\phi \over im\omega} -{\omega_{bz} \over m \omega^2} \cos\theta \sin\phi \right]  \right )  \nonumber \\ 
v_{x\epsilon_2} & = & (1- \omega_b^2/\omega^2-2i\Gamma\omega_b^2/\omega^3)^{-1} \left( eE_{02} \left [ -{\sin\phi \over im\omega} - {\omega_{bz} \over m\omega^2} \cos\phi \right]  \right) \nonumber  \\
v_{y\epsilon_1} & = &  (1- \omega_b^2/\omega^2-2i\Gamma\omega_b^2/\omega^3)^{-1} \left (eE_{01} \left[ {\omega_{by} \omega_{bz} \over im \omega^3} \sin\theta +{\omega_{bz} \over m\omega^2} \cos\phi\cos\theta + \left ( {\omega_{bz}^2 \over i m\omega^3} + {(1-\omega_b^2/\omega^2)\over im \omega} \right) \cos\theta \sin\phi \right ]  \right )  \nonumber \\
v_{y\epsilon_2} & =  & (1- \omega_b^2/\omega^2-2i\Gamma\omega_b^2/\omega^3)^{-1} \left (   e E_{02} \left[  -{\omega_{bz} \over m \omega^2} \sin\phi + \left ({\omega_{bz}^2 \over i m \omega^3} +{(1-\omega_b^2/\omega^2) \over im \omega}\right) \cos\phi \right]  \right)  
\label{eq:finsol1_decay}
\end{eqnarray}
Here we use the following notation for resultant velocities: $v_{x\epsilon_1}$ corresponds to the x-component of the velocity
if the incoming photon is polarized along the $\epsilon_1$ direction, etc.  We do not give the expression for the $z$-component  of the velocity as it plays no role in the computation of Stokes' parameters. 
Here $\omega_{bz} = eB_z/(m c) = e B \cos\alpha / (mc)$, $\omega_{by} = eB_y/(m c) = e B \sin\alpha / (mc)$, and $\omega_b = e \sqrt{B_z^2 + B_y^2}/(m c)$, where $\alpha$ is the angle between $\bm{B}$, and the outgoing photon direction $\hat{z}$. Here, we note that $v_{x} = v_{x\epsilon_1} + v_{x\epsilon_2}$, and  $v_{y} = v_{y\epsilon_1} + v_{y\epsilon_2}$.

\subsection{Polarization calculation - The Stokes Parameters}

The power emitted per solid angle  in a given polarization state, $\mathbf{\epsilon_{fi}}$, is (see e.g. \citealt{1998clel.book.....J} for details):
\begin{equation}
{dP \over d\Omega} = {e^2 \over 4\pi c^3} |\mathbf{\epsilon_{fi}.\dot v}|^2
\label{eq:powpol}
\end{equation}
Here $i = \{1,2\}$

The Stokes' parameters can be defined as follows (see \citealt{1998clel.book.....J} for more details) and computed using  Eq.~(\ref{eq:powpol}): 
\begin{eqnarray}
I & = &  {e^2 \over 4\pi c^3} \left(|\mathbf{\epsilon_{f1}.\dot v}|^2 + |\mathbf{\epsilon_{f2}.\dot v}|^2 \right ) \nonumber \\ 
Q, U, V & = & {e^2 \over 4\pi c^3} \left(|\mathbf{\epsilon_{f1}.\dot v}|^2 - |\mathbf{\epsilon_{f2}.\dot v}|^2 \right )
\end{eqnarray}
Here $\epsilon_{f1} = \hat x$ and $\epsilon_{f2} = \hat y$ for the computation of $I$, and $Q$,  $\epsilon_{f1} = (\hat x + \hat y)/\sqrt{2}$
and $\epsilon_{f2} = (\hat x - \hat y)/\sqrt{2}$ for the computation of U, and $\epsilon_{f1} = (\hat x +i \hat y)/\sqrt{2}$
and $\epsilon_{f2} = (\hat x - i\hat y)/\sqrt{2}$ for the computation of V.  This gives:
\begin{eqnarray}
I & = & (e^2/4 \pi c^3) \omega^2 \Biggl( v_{x} v_{x}^*  + v_{y} v_{y}^* \Biggr) \nonumber \\
Q & = & (e^2/4 \pi c^3) \omega^2 \Biggl( v_{x} v_{x}^*  - v_{y} v_{y}^* \Biggr) \nonumber \\
U & = & (e^2/8 \pi c^3) \omega^2 \Biggl( (v_{x} + v_{y})(v_{x}^* + v_{y}^*)  - (v_{x} - v_{y})(v_{x}^* - v_{y}^*) \Biggr) = (e^2/4 \pi c^3) \omega^2 \Biggl(v_{x} v_{y}^* + v_{y} v_{x}^* \Biggr) \nonumber \\
V & = & (e^2/8 \pi c^3) \omega^2 \Biggl( (v_{x} + iv_{y})(v_{x}^* - iv_{y}^*)  - (v_{x} - iv_{y})(v_{x}^* + iv_{y}^*) \Biggr) = (e^2/4 \pi c^3) \omega^2 \Biggl(i v_{y} v_{x}^* - iv_{x} v_{y}^* \Biggr)
\label{eq:stokes_defn}
\end{eqnarray}
Eqs.~(\ref{eq:stokes_defn}) yield  Stokes parameters for fully polarized incoming light.  A generally elliptically polarized incoming light can be expressed as: ${\bm E} = E_{01} \hat \epsilon_1 + E_{02} \hat \epsilon_2$. This allows us to compute various subcases: for instance,  fully polarized light along $\hat \epsilon_1$ corresponds to $E_{02} =0$ and $E_{01}$ real; circularly polarized light corresponds to $E_{01} = \pm i E_{02}$, etc. For computing the Stokes parameters for unpolarized incoming light, we note that such light could be represented as an incoherent addition of two waves of equal strength orthogonal to each other. So we could compute Stokes parameters for unpolarized incoming light by averaging Stokes parameters for two fully linearly polarized light waves
of equal strength ($E_{01} = E_{02}$) along two orthogonal directions ($\hat \epsilon_1$ and $\hat \epsilon_2$). Similarly, a partially polarized incoming wave corresponds to an average of two waves such that  $E_{01} \ne E_{02}$. Other cases can be worked out using the same logic. From Eqs.~(\ref{eq:stokes_defn}), it can also be verified that for fully polarized light: $I^2 = Q^2+U^2+V^2$ while $I^2 > Q^2+U^2+V^2$ for 
partially polarized light. 

From Eqs.~(\ref{eq:finsol1_decay}) and~(\ref{eq:stokes_defn}), the Stokes parameters can be computed \footnote{\url{https://www.wolfram.com/mathematica/}}: 
\begin{eqnarray}
     I = (e^2/4 \pi c^3) \omega^2 \frac{e^2}{4 m^2\,\left(\omega^6 - 2\omega^4\omega_b^2 + (4\Gamma^2+\omega^2)\omega_b^4\right)}
    \Biggl[
2 E_{02}^*
\Bigl(
E_{02}\left(2\omega^4 - 2\omega^2(\omega_b^2 - 2\omega_{bz}^2) + (\omega_b^2 - \omega_{bz}^2)^2\right)
\nonumber \\
- (2\omega^2 - \omega_b^2 + \omega_{bz}^2)
\Bigl(E_{02}(\omega_b-\omega_{bz})(\omega_b+\omega_{bz})\cos 2\phi
- 2E_{01}\omega_{by}\omega_{bz}\cos\phi\sin\theta\Bigr)
\nonumber \\
+ 2 i E_{01}\omega\omega_{by}(\omega^2+\omega_{bz}^2)\sin\theta\sin\phi
\nonumber \\
+ E_{01}(2\omega^2-\omega_b^2+\omega_{bz}^2)\cos\theta
\Bigl(2 i \omega\omega_{bz} + (-\omega_b+\omega_{bz})(\omega_b+\omega_{bz})\sin 2\phi\Bigr)
\Bigr)
\nonumber \\
+ E_{01}^*
\Bigl(
E_{01}\left(2\omega^4 + \omega_b^4 - 2(\omega_b-\omega_{by})(\omega_b+\omega_{by})\omega_{bz}^2 + \omega_{bz}^4 + 2\omega^2(-\omega_b^2+\omega_{by}^2+2\omega_{bz}^2)\right)
\nonumber \\
+ E_{01}\left(2\omega^4 + \omega_b^4 - 2(\omega_b^2+\omega_{by}^2)\omega_{bz}^2 + \omega_{bz}^4 - 2\omega^2(\omega_b^2+\omega_{by}^2-2\omega_{bz}^2)\right)\cos 2\theta
\nonumber \\
+ 2(2\omega^2-\omega_b^2+\omega_{bz}^2)
\Bigl(E_{01}(\omega_b-\omega_{bz})(\omega_b+\omega_{bz})\cos^2\theta\cos 2\phi
+ 2E_{02}\omega_{by}\omega_{bz}\cos\phi\sin\theta\Bigr)
\nonumber \\
- 4 i E_{02}\omega\omega_{by}(\omega^2+\omega_{bz}^2)\sin\theta\sin\phi
\nonumber \\
+ 2(2\omega^2-\omega_b^2+\omega_{bz}^2)\cos\theta
\Bigl(-2 i E_{02}\omega\omega_{bz}
+ 4E_{01}\omega_{by}\omega_{bz}\sin\theta\sin\phi
\nonumber \\
\qquad\qquad\qquad
+ E_{02}(-\omega_b+\omega_{bz})(\omega_b+\omega_{bz})\sin 2\phi\Bigr)
\Bigr)
\Biggr] \nonumber
\end{eqnarray}

\begin{eqnarray}
     Q = -(e^2/4 \pi c^3) \omega^2 \frac{e^2}{4 m^2\,\left(\omega^6 - 2\omega^4\omega_b^2 + (4\Gamma^2+\omega^2)\omega_b^4\right)}
    \Biggl[
2 E_{02}^*
\Bigl(
E_{02}\left(2\omega^4 - 2\omega^2 \omega_b^2 + (\omega_b^2 - \omega_{bz}^2)^2\right)\cos 2\phi
\nonumber \\
+ (\omega_b-\omega_{bz})(\omega_b+\omega_{bz})
\Bigl(E_{02}(-2\omega^2 + \omega_b^2 - \omega_{bz}^2)
- 2E_{01}\omega_{by}\omega_{bz}\cos\phi\sin\theta\Bigr)
\nonumber \\
+ 2 i E_{01}\omega\omega_{by}(-\omega+\omega_{bz})(\omega+\omega_{bz})\sin\theta\sin\phi
\nonumber \\
+ E_{01}\cos\theta
\Bigl(2 i \omega\omega_{bz}(-\omega_b+\omega_{bz})(\omega_b+\omega_{bz})
+ (2\omega^4 - 2\omega^2 \omega_b^2 + (\omega_b^2-\omega_{bz}^2)^2)\sin 2\phi\Bigr)
\Bigr)
\nonumber \\
+ E_{01}^*
\Bigl(
E_{01}\Bigl(\omega_b^4 - 2(\omega_b-\omega_{by})(\omega_b+\omega_{by})\omega_{bz}^2 + \omega_{bz}^4
- 2\omega^2(\omega_b^2+\omega_{by}^2-\omega_{bz}^2)
\nonumber \\
\qquad\quad
+ (\omega_b^4 - 2(\omega_b^2+\omega_{by}^2)\omega_{bz}^2 + \omega_{bz}^4
+ 2\omega^2(-\omega_b^2+\omega_{by}^2+\omega_{bz}^2))\cos 2\theta\Bigr)
\nonumber \\
- 2 E_{01}(2\omega^4 - 2\omega^2 \omega_b^2 + (\omega_b^2-\omega_{bz}^2)^2)\cos^2\theta \cos 2\phi
\nonumber \\
+ 4 E_{02}\omega_{by}\omega_{bz}(-\omega_b+\omega_{bz})(\omega_b+\omega_{bz})\cos\phi\sin\theta
\nonumber \\
+ 4 i E_{02}\omega\omega_{by}(\omega-\omega_{bz})(\omega+\omega_{bz})\sin\theta\sin\phi
\nonumber \\
+ 2\cos\theta
\Bigl(
2\omega_{bz}(-\omega_b+\omega_{bz})(\omega_b+\omega_{bz})
(- i E_{02}\omega + 2E_{01}\omega_{by}\sin\theta\sin\phi)
\nonumber \\
\qquad\quad
+ E_{02}(2\omega^4 - 2\omega^2 \omega_b^2 + (\omega_b^2-\omega_{bz}^2)^2)\sin 2\phi
\Bigr)
\Bigr)
\Biggr] \nonumber
\end{eqnarray}

\begin{eqnarray}
     U = (e^2/4 \pi c^3) \omega^2 \frac{e^2\,\omega(\omega-\omega_b)(\omega+\omega_b)}{m^2\,\left(\omega^6 - 2\omega^4\omega_b^2 + (4\Gamma^2+\omega^2)\omega_b^4\right)}
    \Biggl[
E_{02}^*
\Bigl(
E_{01}\,\omega\cos\theta\cos 2\phi
- \cos\phi\Bigl(i E_{01}\omega_{by}\sin\theta + 2E_{02}\omega\sin\phi\Bigr)
\Bigr)
\nonumber \\
+ E_{01}^*
\Bigl(
E_{02}\,\omega\cos\theta\cos 2\phi
+ i E_{02}\omega_{by}\cos\phi\sin\theta
+ E_{01}\,\omega\cos^2\theta\sin 2\phi
\Bigr)
\Biggr] \nonumber
\end{eqnarray}

\begin{eqnarray}
     V = (e^2/4 \pi c^3) \omega^2 \frac{e^2\,\omega}{2 m^2\,\left(\omega^6 - 2\omega^4\omega_b^2 + (4\Gamma^2+\omega^2)\omega_b^4\right)}
    \Biggl[
-2 E_{02}^*
\Bigl(
E_{02}\omega_{bz}\left(2\omega^2 - \omega_b^2 + \omega_{bz}^2 + (-\omega_b^2 + \omega_{bz}^2)\cos 2\phi\right)
\nonumber \\
+ E_{01}\omega_{by}\sin\theta
\Bigl((\omega^2 - \omega_b^2 + 2\omega_{bz}^2)\cos\phi + 2 i \omega \omega_{bz}\sin\phi\Bigr)
\nonumber \\
+ E_{01}\cos\theta
\Bigl(i \omega (\omega^2 - \omega_b^2 + 2\omega_{bz}^2)
+ \omega_{bz}(-\omega_b+\omega_{bz})(\omega_b+\omega_{bz})\sin 2\phi\Bigr)
\Bigr)
\nonumber \\
- E_{01}^*
\Bigl(
E_{01}\omega_{bz}\left(2\omega^2 - \omega_b^2 + 2\omega_{by}^2 + \omega_{bz}^2 
+ (2\omega^2 - \omega_b^2 - 2\omega_{by}^2 + \omega_{bz}^2)\cos 2\theta\right)
\nonumber \\
+ 2E_{01}(\omega_b-\omega_{bz})\omega_{bz}(\omega_b+\omega_{bz})\cos^2\theta\cos 2\phi
\nonumber \\
+ 2E_{02}\omega_{by}(\omega^2 - \omega_b^2 + 2\omega_{bz}^2)\cos\phi\sin\theta
\nonumber \\
- 4 i E_{02}\omega\omega_{by}\omega_{bz}\sin\theta\sin\phi
\nonumber \\
+ 2\cos\theta
\Bigl(
(\omega^2 - \omega_b^2 + 2\omega_{bz}^2)(- i E_{02}\omega + 2E_{01}\omega_{by}\sin\theta\sin\phi)
\nonumber \\
\qquad\quad
+ E_{02}\omega_{bz}(-\omega_b+\omega_{bz})(\omega_b+\omega_{bz})\sin 2\phi
\Bigr)
\Bigr)
\Biggr] 
\label{eq:stokespar_sol}
\end{eqnarray}

As noted above, these are Stokes parameters for a fully polarized light. We need to average Stokes parameters over time to convert them into observables. For a fully polarized source, this averaging is trivial as $E_{01}$ and $E_{02}$ are constants.  For an unpolarized
incoming source, $\langle E_{01} E_{02} \rangle = 0$ with $E_{01} = E_{02}$. The linearly partially polarized case is best treated by assuming $\langle E_{01} E_{02} \rangle = 0$ but $E_{01} \ne E_{02}$ (for details of the representation of partially polarized light, see e.g., 
\citealt{1971ctf..book.....L}, section 50). 
The Stokes parameters depend on the direction of incoming photons, i.e., $\theta$, and $\phi$, the magnetic field strength $B$, and the angle between $B$ and the outgoing direction $\hat{z}$, i.e., $\alpha$. 

The expressions for Stokes parameters in Eqs.~(\ref{eq:stokespar_sol})  are long and cumbersome (see \citealt{1986Ap&SS.121..333C} for an alternative treatment and references therein). So we discuss some simple,
well-known solutions of the physical setting. 

First,  it can be verified that $V = 0$ if the magnetic field is zero, as Thompson scattering of unpolarized photons off charged particles only
generates linearly polarized light (e.g., \citealt{1986rpa..book.....R} for details). From Eqs.~(\ref{eq:finsol1_decay}) we can show
in general that circular polarization vanishes if the velocities are either purely real or imaginary. The circular polarization 
is only generated if the velocity components are complex, and the degree of circular polarization is determined by the relative 
strengths of the real and imaginary components. This allows us to reach many other inferences which would be relevant in 
interpreting our results for incoming unpolarized light:  (i) if $\omega \gg \omega_b$, the scattered light is predominantly linearly polarized, (ii) if  $\omega_b \gg \omega$, the circular polarization becomes negligible. In both these cases, the circular polarization 
is small as the velocity components are dominated by either the real or the imaginary component.  Another result, which follows
from physical intuition, corresponds to the case where the incoming unpolarized photons arrive from the $z$-direction ($\theta = \phi = 0$) and the $z$-component of the magnetic field vanishes. In this case, the circular polarization vanishes. 

To get further insight into the physical interpretation of Stokes parameters,  we next consider the following case: an incoming photon in the $z$-direction and a magnetic field in the $z$-direction ($B_y =0$). Further assuming $\Gamma = 0$, we get:
\begin{eqnarray}
v_x & =  &  \left[ {e E_{01}  \over im \omega} - {e E_{02} \omega_b \over m \omega^2} \right] \times (1- \omega_b^2/\omega^2)^{-1} \nonumber \\
v_y & = &   \left[ {e E_{02}  \over im \omega} + {e E_{01} \omega_b \over m \omega^2} \right] \times (1- \omega_b^2/\omega^2)^{-1}  \nonumber \\
v_z & =  & 0
\label{eq:finsolsimp}
\end{eqnarray}

This can be used to compute the response of the electron to right (left) handed (R(L)H)  circularly polarized light.  In this
case, $E_{01} = \pm i E_{02}$, which gives:
\begin{eqnarray}
v_x & =  &  {eE_{01}  \over i m\omega}\left (1  \pm {\omega_{b} \over \omega}  \right)\times (1- \omega_b^2/\omega^2)^{-1} \nonumber \\
v_y & = &  {eE_{01}  \over m\omega}\left ( {\omega_{b} \over \omega}  \pm 1 \right)\times (1- \omega_b^2/\omega^2)^{-1} \nonumber \\
v_z & =  & 0
\label{eq:finsolz}
\end{eqnarray}
This case shows that the response of the electron to  RH(LH) polarized light velocity scales as $1/(1\pm(\omega_b/\omega))$. 
The cross-section for scattering can be obtained by dividing the relevant expressions by the incoming flux, $c (E_{01}^2 + E_{02}^2)/(8\pi)$. This proves the well known result that  for $\omega_b \simeq \omega$ 
the cross-section of scattering is enhanced(suppressed) for LH(RH) circularly polarized light (e.g. \citealt{1971PhRvD...3.2303C}). More generally, it follows from Eqs.~(\ref{eq:stokespar_sol}) that  the cross section for light scattering close to the  resonance sharply increases, by roughly a factor of $(\omega_b/\Gamma)^2$

To make further progress, we assume physical parameters of neutron stars. The surface magnetic field
for normal pulsars, $B \simeq 10^{12} \, \rm G$. For a magnetic field of this strength,  the cross-section of scattering increases by 
a factor of nearly $10^7$ close to the resonance frequency. 

The incoming photons originate in the accreting column close to the surface of the neutron stars and, depending on  the number density of non-relativistic electrons and the column height, could scatter multiple times or just once before reaching the observer. The height and the width of the accreting column are  expected to be around 10~kms and  1~km, respectively. The number density of non-relativistic electrons in the accretion column and close to  the surface of neutron stars can vary substantially. In accreting neutron stars, the density could be in the range $10^{21} \hbox{--} 10^{23} \, \rm cm^{-3}$.  For magnetars, assuming them to be isolated systems, the number density would be guided by the Goldreich-Julian limit \citep{Goldreich} and the condition of pair cascade, and could lie in the range  $10^{13} \hbox{--} 10^{18} \, \rm cm^{-3}$. These numbers can also vary for different scenarios within single source, e.g., burst or flare. This scenario is further complicated by the strong energy dependence of the Thompson cross-section close to the resonance frequency.  Given such a diversity of physical parameters, we consider two cases to capture most of the relevant outcomes: (a) the medium is optically thick, and the photon that originates in an accretion column scatters multiple times before escaping. In this case, the direction of the incoming photons could be assumed to lie in the entire hemisphere before scattering in the forward direction, and (b) the photon originates close to the neutron star surface, e.g., magnetars, and scatters just once before reaching the observers, i.e., optically thin regime.

\subsection{Multiple scattering}
If the mean free path of the photons is short and they scatter multiple times, their direction of arrival before the final
scattering is not related to the point of origin. Even though it is hard to model this situation owing to the complications of the geometry for  different NS systems, we mimic this situation by assuming that the photons arise  from the entire backward hemisphere before the final scattering.
Normally, we expect the photons to become unpolarized after multiple scatterings, but we also study incoming light of arbitrary polarization.
This caters to many intermediate situations, e.g., a few scatterings, and  the possibility of one of the modes of polarization
being scattered many times as compared to the other, e.g., the discussion leading to Eq.~(\ref{eq:finsolz}). 
Therefore, to study the case where the medium is   optically thick,  we obtain the angle-integrated Stokes parameters: 
\begin{eqnarray}
\int_{0}^{\pi/2}\int_{0}^{2\pi} I \sin\theta \, d\phi \, d\theta &=& 
(e^2/4 \pi c^3) \omega^2 \frac{e^2\,\pi}{3 m^2\,\left(\omega^6 - 2\omega^4\omega_b^2 + (4\Gamma^2+\omega^2)\omega_b^4\right)}
\Biggl[
\Bigl(
-3 i E_{02}\,\omega\,\omega_{bz}\left(2\omega^2 - \omega_b^2 + \omega_{bz}^2\right)
\nonumber \\
&& \quad {}+ E_{01}\Bigl(
2\omega^4 + \omega_b^4 - 2(\omega_b^2 - 2\omega_{by}^2)\omega_{bz}^2 + \omega_{bz}^4
+ \omega^2(-2\omega_b^2 + 4(\omega_{by}^2 + \omega_{bz}^2))
\Bigr)
\Bigr) E_{01}^*
\nonumber \\
&& \quad {}+ 3 \Bigl(
i E_{01}\,\omega\,\omega_{bz}\left(2\omega^2 - \omega_b^2 + \omega_{bz}^2\right)
\nonumber \\
&& \quad \quad {}+ E_{02}\Bigl(
2\omega^4 - 2\omega^2(\omega_b^2 - 2\omega_{bz}^2)
+ (\omega_b^2 - \omega_{bz}^2)^2
\Bigr)
\Bigr) E_{02}^*
\Biggr] \nonumber
\end{eqnarray}

\begin{eqnarray}
\int_{0}^{\pi/2}\int_{0}^{2\pi} Q \sin\theta \, d\phi \, d\theta &=& 
(e^2/4 \pi c^3) \omega^2 \frac{e^2\,\pi}{3 m^2\,\left(\omega^6 - 2\omega^4\omega_b^2 + (4\Gamma^2+\omega^2)\omega_b^4\right)}
\Biggl[
-\Bigl(
3 i E_{02}\,\omega\,(\omega_b-\omega_{bz})\omega_{bz}(\omega_b+\omega_{bz})
\nonumber \\
&& \quad {}+ E_{01}\Bigl(
\omega_b^4 - 2(\omega_b^2 - 2\omega_{by}^2)\omega_{bz}^2 + \omega_{bz}^4
- 2\omega^2(\omega_b^2 + 2\omega_{by}^2 - \omega_{bz}^2)
\Bigr)
\Bigr) E_{01}^*
\nonumber \\
&& \quad {}- 3(\omega_b-\omega_{bz})(\omega_b+\omega_{bz})
\Bigl(
- i E_{01}\,\omega\,\omega_{bz}
+ E_{02}(-2\omega^2 + \omega_b^2 - \omega_{bz}^2)
\Bigr) E_{02}^*
\Biggr] \nonumber
\end{eqnarray}

\begin{eqnarray}
\int_{0}^{\pi/2}\int_{0}^{2\pi} U \sin\theta \, d\phi \, d\theta &=& 
0 \nonumber
\end{eqnarray}

\begin{eqnarray}
\int_{0}^{\pi/2}\int_{0}^{2\pi} V \sin\theta \, d\phi \, d\theta &=& 
(e^2/4 \pi c^3) \omega^2 \frac{e^2\,\pi\,\omega}{3 m^2\,\left(\omega^6 - 2\omega^4\omega_b^2 + (4\Gamma^2+\omega^2)\omega_b^4\right)}
\Biggl[
-2 E_{01}\,\omega_{bz}\left(2\omega^2 - \omega_b^2 + 4\omega_{by}^2 + \omega_{bz}^2\right) E_{01}^*
\nonumber \\
&& \quad {}+ 3 i E_{02}\,\omega\left(\omega^2 - \omega_b^2 + 2\omega_{bz}^2\right) E_{01}^*
\nonumber \\
&& \quad {}- 6 E_{02}\,\omega_{bz}\left(2\omega^2 - \omega_b^2 + \omega_{bz}^2\right) E_{02}^*
\nonumber \\
&& \quad {}- 3 i E_{01}\,\omega\left(\omega^2 - \omega_b^2 + 2\omega_{bz}^2\right) E_{02}^*
\Biggr]
\label{eq:mul_scat_Stokes}
\end{eqnarray}

We have already discussed some generic features of Stokes parameters of scattered photons in the presence of a magnetic field.
Those are still applicable to angle-integrated Stokes parameters, but the angle-integrated Stokes parameters introduce additional
features. For instance, in case of unpolarized incoming light, for  $B=0$, all the Stokes parameters vanish in this case\footnote{This is expected for the following reasons: for $B=0$, $\{Q,U\} \propto \{\sin(2\phi),\cos(2\phi)\}$, which vanish upon integration. $V=0$, as noted above, for zero magnetic field. \label{fn:quwo}}. Therefore, this case provides 
a good baseline model for isolating the impact of a high magnetic field on the observed polarized light. 

\begin{figure}[h]
    \centering
    \includegraphics[width=\linewidth, keepaspectratio]{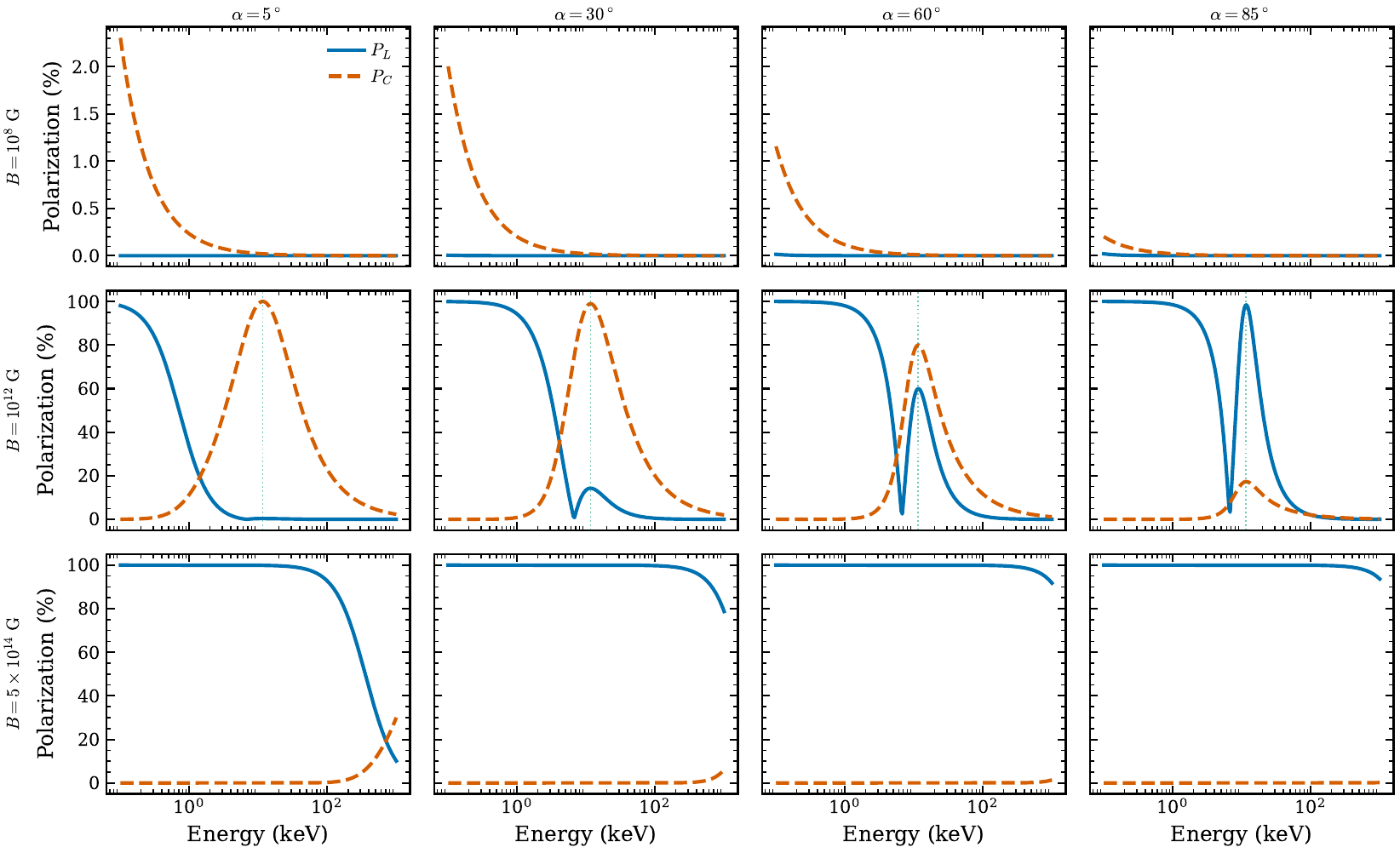} \\
    \caption{Linear ($P_L = \sqrt{Q^2+U^2}/ I$) and circular polarization ($P_C = |V|/I$) of scattered photons are shown as a function of photon energy for different outgoing angles for integrated incoming angles and a fixed magnetic field, i.e.,  $10^{8}$~G, $10^{12}$~G and $5 \times 10^{14}~G$ (top to bottom respectively) for incoming unpolarized light. }
    \label{fig:Stokes_unpolarized_incoming}
\end{figure}

In Figure~\ref{fig:Stokes_unpolarized_incoming}, we show the expected degree of polarization  (Eq.~(\ref{eq:mul_scat_Stokes})) for unpolarized incoming photons as a function of energy for  three magnetic field strengths: $B = \{10^8, 10^{12}, 5 \times 10^{14}\} \, \rm G$
(the gyration frequency, $\nu_b \equiv \omega_b/(2\pi) =  11.6 (B/10^{12} \, \rm G) \, \rm keV$) and four angles between the outgoing photon and the magnetic field. 

For the smallest magnetic field, $\omega \gg \omega_b$, the entire range of energies is shown in the figure.  This could correspond to low-magnetized neutron stars like LMXBs. In this limit, we have argued above that, before the angle integration, the scattered light is expected to be mostly
linearly polarized. After angle integration, the linear polarization is driven close to  zero  for a small  magnetic field,  while the circular polarization is also  small because 
$\omega \gg \omega_b$ (Footnote~\ref{fn:quwo}).

For $B = 10^{12} \, \rm G$, a more realistic value on the surface of normal pulsars, the polarization pattern is more 
complex. As expected on the basis of our earlier discussion, the circular polarization peaks at the resonance frequency, $\omega \simeq \omega_b$.  One of the striking features of this case is that the scattered photons could be strongly circularly polarized
for smaller $\alpha$, the angle between the outgoing photon and the magnetic field.   Linear polarization also shows 
interesting patterns close to the resonance frequency: $Q$ reaches zero and changes sign close to the resonance frequency, and the degree of linear polarization also peaks at the resonance frequency. 

The final case, $B = 5 \times 10^{14} \, \rm G$, is the expected value on the surface of the magnetars. It 
captures the situation $\omega \ll \omega_b$ for the energy range shown in the figure. As for the case 
$\omega \gg \omega_b$, we expect the circular polarization to be negligible in this case also. However, a marked difference between the polarization structure between this case and a small magnetic field is the degree of linear polarization. When the magnetic field
is negligible, the angle-integrated linear polarization vanishes, as seen in the top panel of this figure. However, 
the linear polarization is very strong for large magnetic field. This striking contrast arises from the difference
in the angular dependence of Stokes parameters in the two cases. 

In sum: for normal pulsars with a magnetic field $B \simeq 10^{12}$~G, the circular polarization dominates over linear polarization over a large range of outgoing angles close to the resonance frequency.  However, for a magnetar-like field strength, i.e., $B \simeq 10^{14}$~G,  the scattered light is expected to be linearly polarized with a high degree of polarization. These results could potentially  explain  the observational properties:  the observed linear polarization degree is low in normal pulsars but high in magnetars. We return to a more detailed discussion comparing our results with observations below. 

\begin{figure}[h]
    \centering
    \includegraphics[width=\linewidth, keepaspectratio]{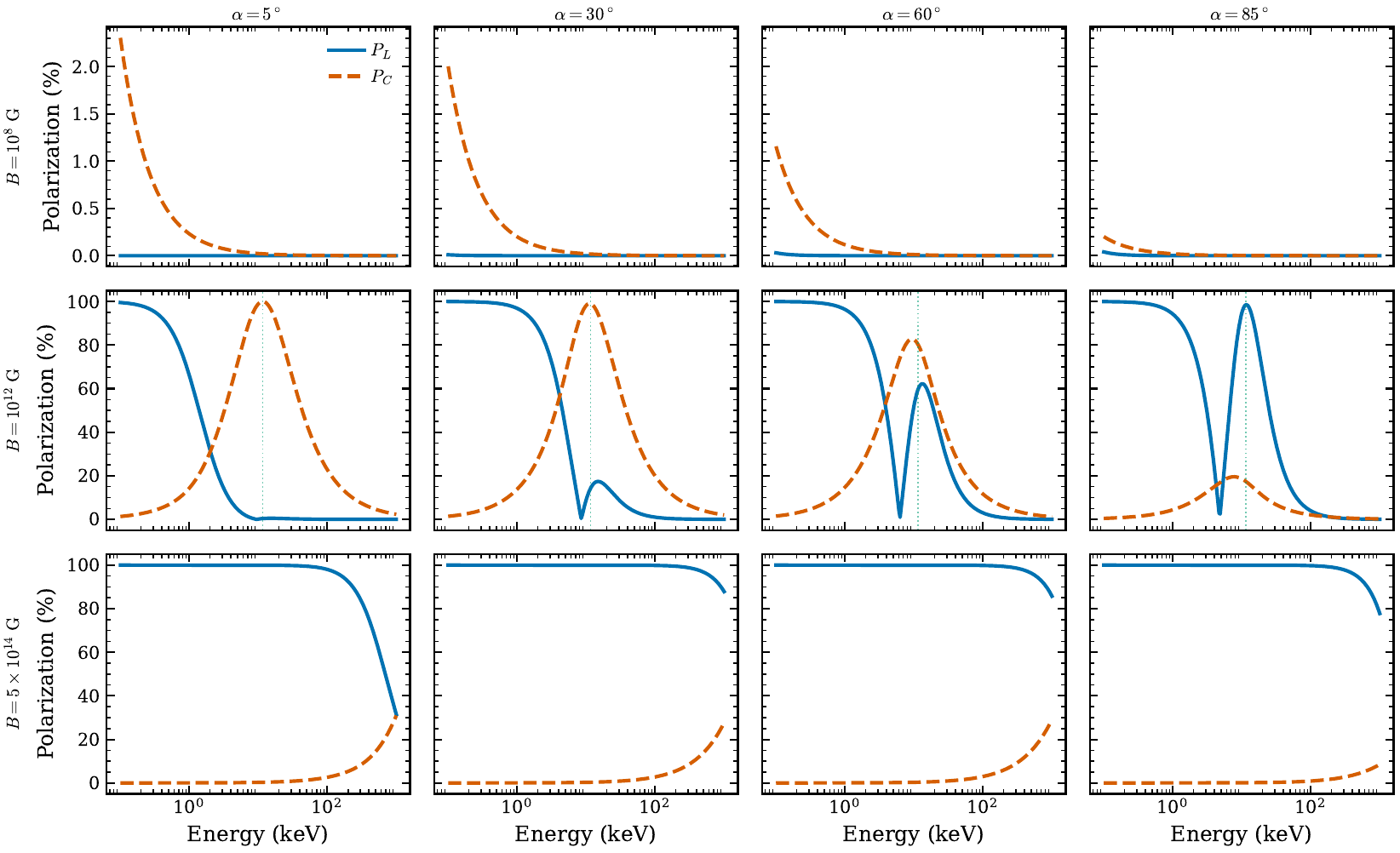} \\
    \caption{Same as Figure~\ref{fig:Stokes_unpolarized_incoming} for the incoming photon to be fully linearly polarized.} 
    \label{fig:Stokes_linear_incoming}
\end{figure}

\begin{figure}[h]
    \centering
    \includegraphics[width=\linewidth, keepaspectratio]{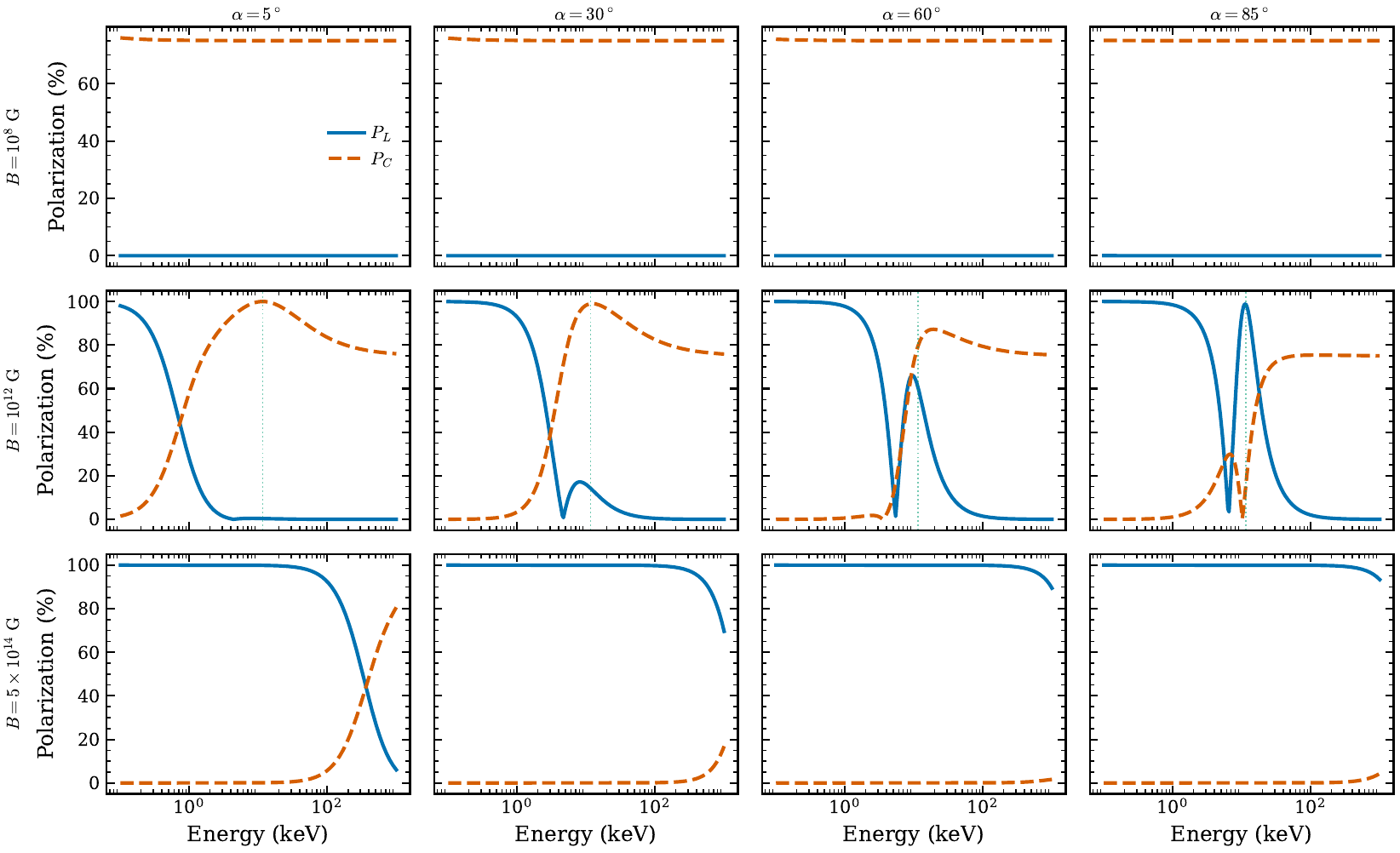} \\
    \caption{Same as Figure~\ref{fig:Stokes_unpolarized_incoming} for the incoming photon to be fully circularly polarized.}
    \label{fig:Stokes_circular_incoming}
\end{figure}

In Figures~\ref{fig:Stokes_linear_incoming} and~\ref{fig:Stokes_circular_incoming}, we show the polarization of scattered 
photons if the incoming photons were fully linearly or circularly polarized. The aim of these figures is to take into account
a possible plethora of initial conditions. For instance, as already discussed above, it is possible under certain geometries
for the cross-section of scattering of right and left-handed polarized lights to be very different, which might mean
the incoming light could be strongly circularly polarized. Similarly, the incoming light could be strongly linearly polarized under
other possible geometries. In addition to cases shown in Figures~\ref{fig:Stokes_linear_incoming} and~\ref{fig:Stokes_circular_incoming},
we have also considered partially linearly and circularly polarized incoming light in our analysis (not shown).

There are some salient differences in the polarization state of the scattered photons for different cases. 
For  $\omega>>\omega_b$,  $|V|/I $ is negligible for incoming linear/unpolarized light, but for circularly polarized incoming light, $|V|/I = 3/4$, irrespective of the outgoing angles. Thus, if the incoming photons are circularly polarized, 
the outgoing photons are also strongly circularly polarized. 

For normal pulsars, where $\omega \simeq \omega_b$, the mode conversion between linear and circular polarization happens close to the resonance frequencies for most non-zero angles, and is nearly independent of the polarization of incoming photons. For the pencil beam case, i.e., $\alpha \approx 0^{\circ}$, the outgoing photons are purely circularly polarized, and for the strong fan beam scenario ($\alpha \approx 90^{\circ}$), the outgoing photons are purely linearly polarized. These dependencies of angles can provide pointers to why we observe lower degrees of polarization in accreting X-ray pulsars.

For $\omega \ll \omega_b$, the linear polarization of outgoing photons is very high and nearly independent of the polarization state of incoming light. 

Even though the initial
conditions have a bearing on the nature of scattered light, it is striking that some of the features remain the same. For instance,
the features close to the resonance frequency for intermediate magnetic field strength are nearly the same. This suggests
some of our results are generic and independent of the initial conditions. We have verified that this remains the case for 
other initial conditions as well.

\subsection{Optically thin case}
In the previous subsection, we considered the case of multiple scattering of photons that isotropizes the photon 
distribution before last scattering. However, it is also conceivable that the region around the source of emission 
is optically thin to photons, e.g., for magnetars, isolated neutron stars. Such isolated systems are likely to have electron number densities much lower than in accreting systems. 
To model this case,  Eqs.~(\ref{eq:stokespar_sol}) yield the requisite solutions of photon polarization as a function of the incoming angles.

\begin{figure}[h]
    \centering
    \includegraphics[width=\linewidth, keepaspectratio]{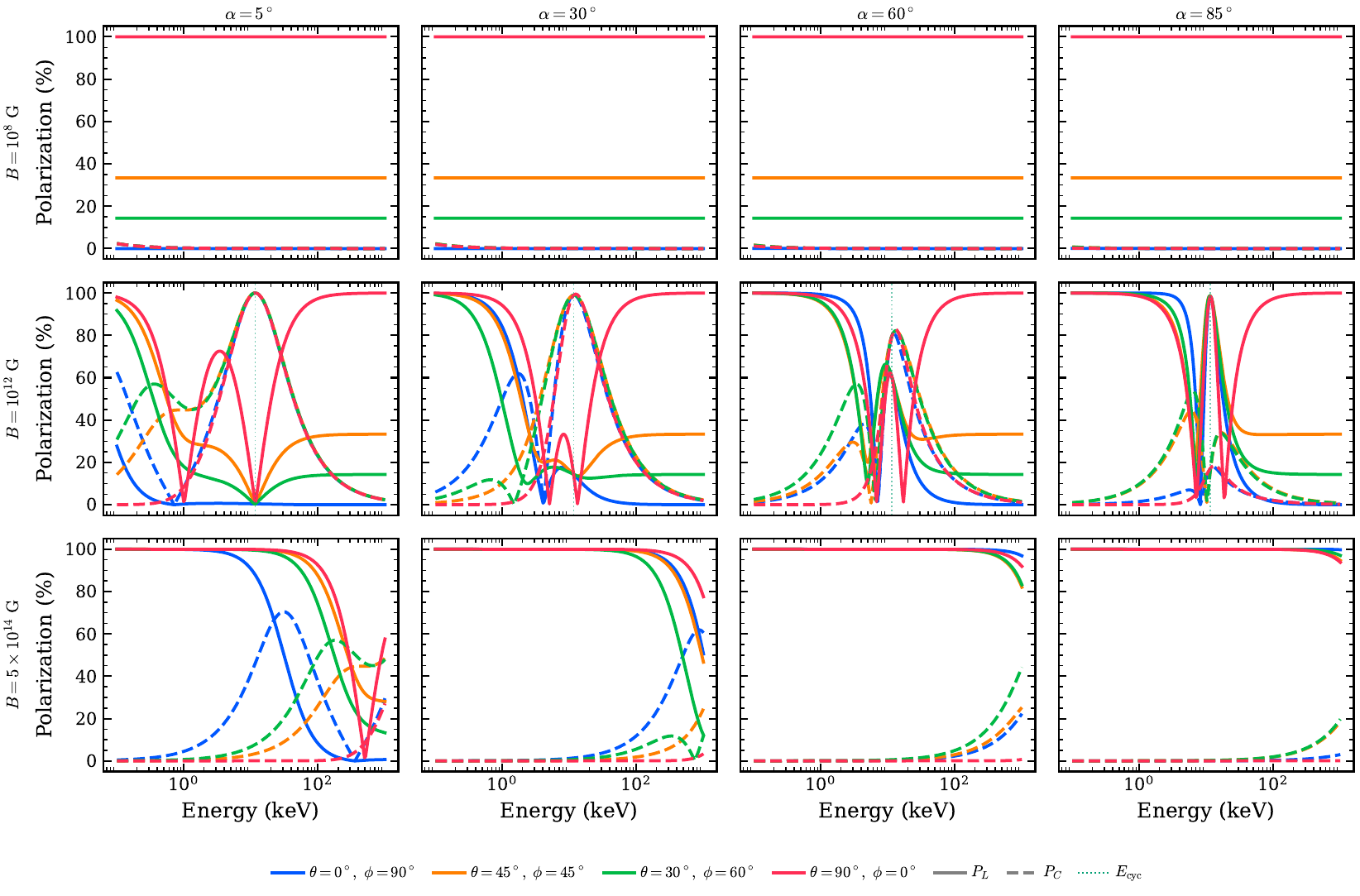} \\
    \caption{Follows the same convention as  Figure~\ref{fig:Stokes_unpolarized_incoming} except for some combinations of incoming angles $\theta$, and $\phi$ (i.e., representing optically thin cases), and for the incoming photons to be unpolarized.}
    \label{fig:Stokes_unpolarized_incoming_bothangles}
\end{figure}

\begin{figure}[h]
    \centering
    \includegraphics[width=\linewidth, keepaspectratio]{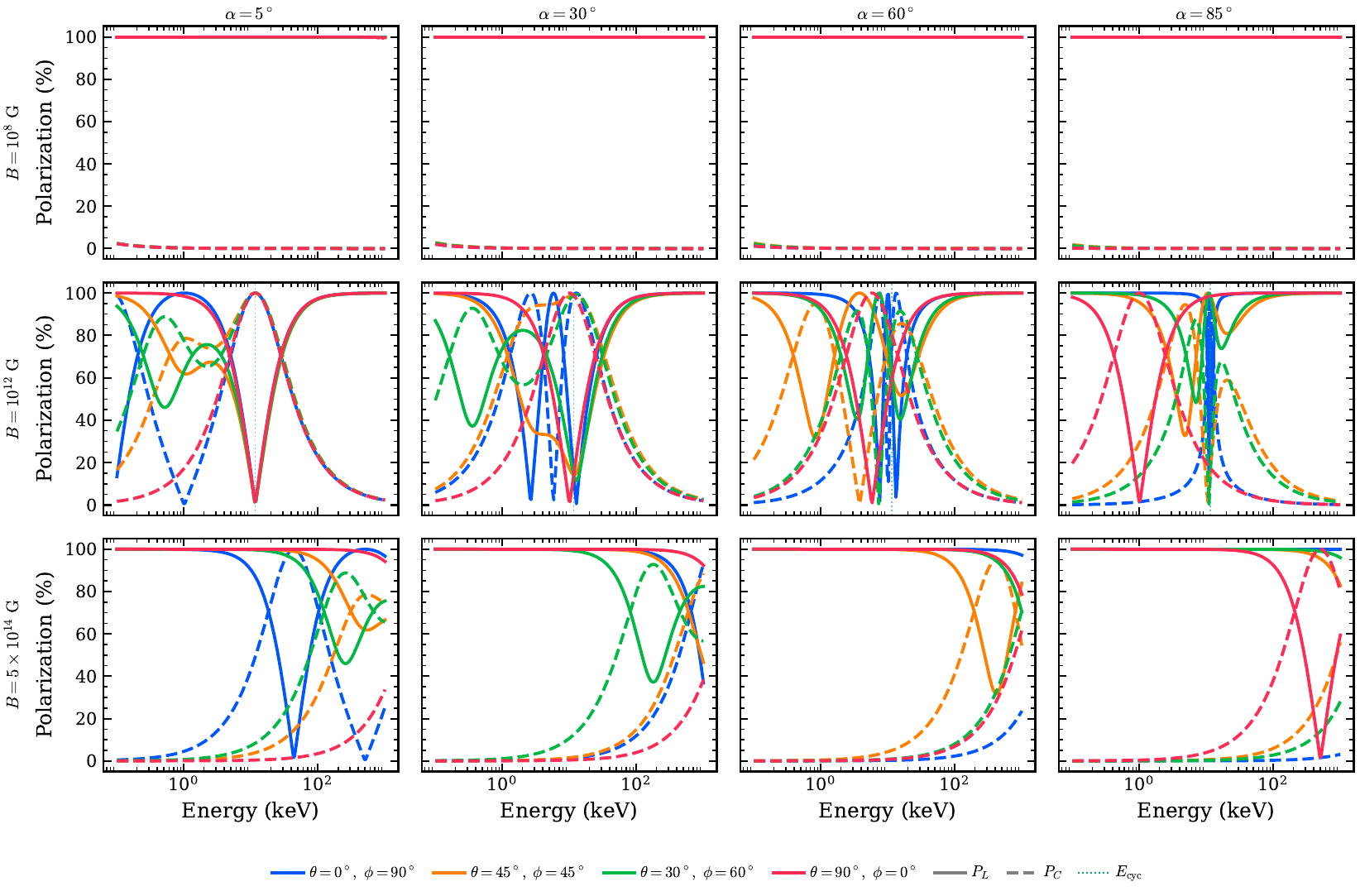} \\
    \caption{Follows the same convention as  Figure~\ref{fig:Stokes_unpolarized_incoming_bothangles} for the incoming photons to be linearly polarized.}
    \label{fig:Stokes_linear_incoming_bothangles}
\end{figure}

\begin{figure}[h]
    \centering
    \includegraphics[width=\linewidth, keepaspectratio]{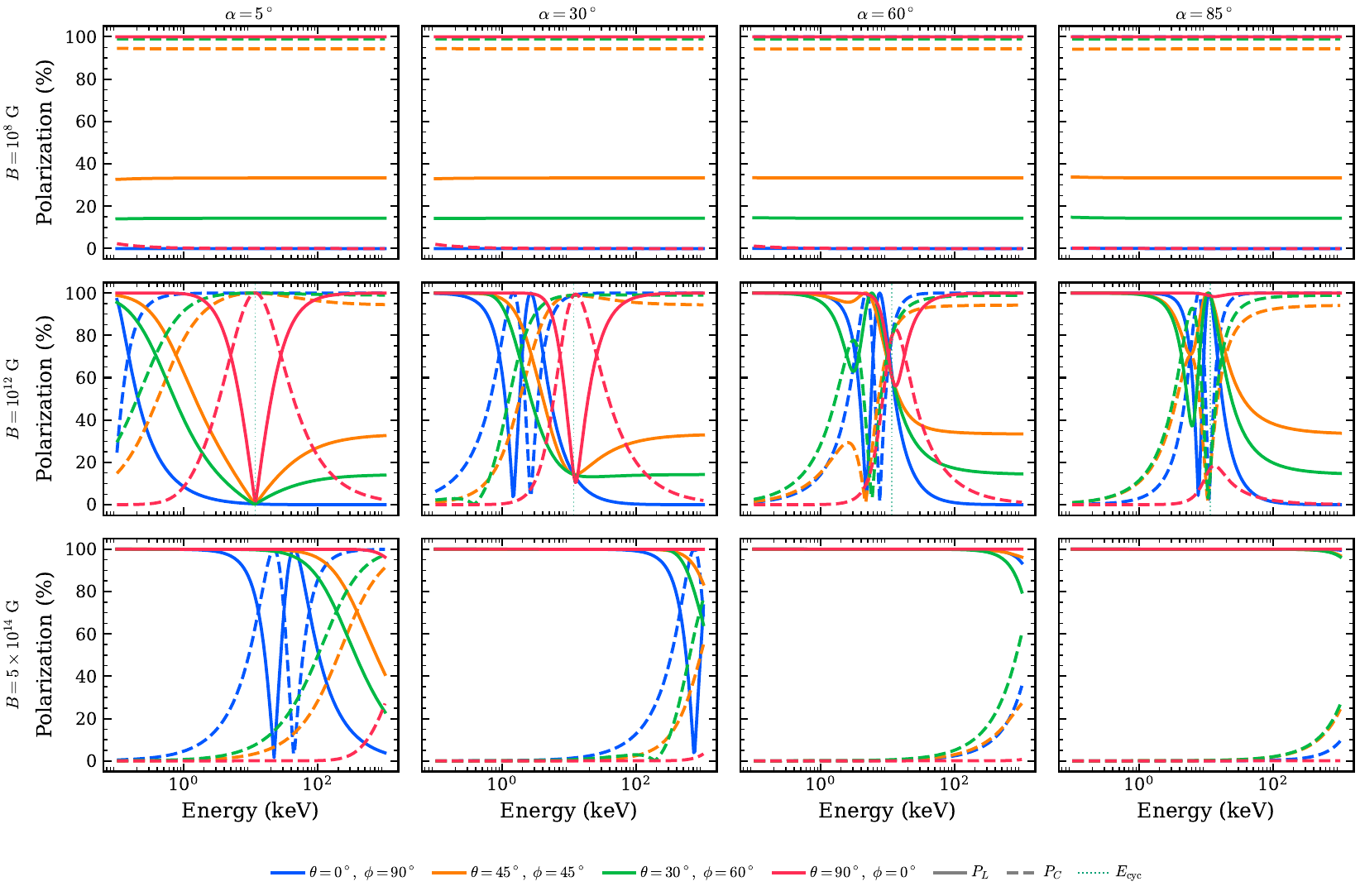} \\
    \caption{Follows the same convention as  Figure~\ref{fig:Stokes_unpolarized_incoming_bothangles} for the incoming photons to be circularly polarized.}
    \label{fig:Stokes_circular_incoming_bothangles}
\end{figure}

In Figure~\ref{fig:Stokes_unpolarized_incoming_bothangles},\ref{fig:Stokes_linear_incoming_bothangles}, and \ref{fig:Stokes_circular_incoming_bothangles}, we display the polarization of the scattered light for different combinations of $\theta$ and $\phi$,   $B$ and $\alpha$. The expanded parameter space introduces additional complexity in the 
polarization of the scattered light. We note some salient differences and similarities between this case and the optically thick case.
\begin{itemize}
\item[(a)] For a low magnetic field, the observed polarization is close to zero for the optically thick case for reasons discussed above. However, in the optically thin case, the observed polarization is non-zero and strongly depends on the incoming angle. 
\item[(b)] For magnetars, the linear polarization dominates as in the previous cases, but it shows more variation with incoming angles. However, the degree of linear polarization remains large. It also falls as energy increases in most cases, while, as discussed below, the observed polarization increases with energy for magnetars. One possible way to explain the data is that the medium around magnetars is optically thin, and scattered photons at different energies arrive from a range of incoming angles.  This means that the spectral dependence of the observed polarization arises from a mix of incoming angles. Given the complicated spectral dependence on the incoming angle (Figure~\ref{fig:Stokes_unpolarized_incoming_bothangles}, \ref{fig:Stokes_linear_incoming_bothangles}, \ref{fig:Stokes_circular_incoming_bothangles}), this might provide a plausible explanation. To compare this theoretical proposal with the data would require the construction of a detailed model
of the source emission,  scattering geometry, and magnetic field. We do not attempt it here. 
\end{itemize}
We note that the optically thin case retains many generic features of the earlier case. 

Before reaching the observer, the scattered photons pass through regions of pervasive plasma and strong magnetic fields. We next seek to establish how the polarization of scattered light is impacted by vacuum and plasma polarization effects.

\section{Vacuum birefringence} \label{sec:vacuum_birefringence}
The propagation of polarized photons in the vicinity of highly magnetized neutron stars is impacted by the vacuum and plasma
birefringence (see e.g. \citealt{Gnedin1978,Heyl2018,2003ApJ...588..962L,2003PhRvL..91g1101L,Sokolova-Lapa2023,Meszaros1988,Caiazzo2021,1983JOSA...73.1719K, 1997JPhA...30.6485H}). 
It is known that in the presence of strong magnetic fields, the electromagnetic vacuum gets polarized. This modifies the 
dielectric constant to (e.g., \citealt{1997JPhA...30.6485H} and references therein): 
\begin{equation}
\epsilon_{ij} = \delta_{ij}  + {\alpha\over 2 \pi} \left[\left (-2 X_0(1/\xi) +{1\over \xi} X_0^{(1)}(1/\xi) \right) \delta_{ij} - {1\over \xi^2} {B_i B_j \over B_{\rm crit}^2}  X_1(1/\xi) \right]
\label{eq:epsarb}
\end{equation}
Here $i, j=1,2$ and $\alpha = 1/137$ is the fine structure constant. Here $B_{\rm crit} = m_e^2 c^3/(e\hbar) = 4.4 \times 10^{13} \, \rm G$ and  $\xi = B/B_{\rm crit}$. 
$X_0$, $X_0^{(1)}$ and $X_1$ can only be expressed in integral forms. We do not list them here; for details
see Equations~(11)--(13) of \citet{1997JPhA...30.6485H}. 
For $B \ll B_{\rm crit}$, Eq.~(\ref{eq:epsarb}) simplifies to:
\begin{equation}
\epsilon_{ij} = \delta_{ij}  + {\alpha\over 45 \pi B_{\rm crit}^2}\left(-2 B^2 \delta_{ij} + 7 B_i B_j \right)
\label{eq:epsanis}
\end{equation}
 We note that the modified dielectric tensor is real and symmetric. 

The plasma effects can be included by the following relation between ${\bf D}$ and ${\bf E}$ (see \citealt{1970pewp.book.....G} for details):. 
\begin{equation}
    {\bf D} = \left(1- {\omega_0^2 \over (\omega^2 - \omega_b^2)} \right){\bf E} + \left ({\omega_0^2 \over \omega^2 - \omega_b^2} \right) {\bm{\omega_b} ({\bf E}.\bm{\omega_b}) \over \omega^2} -i \left ({\omega_0^2 \over (\omega^2 - \omega_b^2)\omega} \right) ({\bf E x}\bm{\omega_b})
\end{equation}
Here $\omega_0$ is the plasma frequency; $\omega_0^2 = 4\pi e^2 n_e/m_e$ and  $\bm{\omega_b} = e {\bf B}/(m_e c)$.  The anisotropic dielectric tensor can be read off from the relation: $D_i = \epsilon_{ij}E_j$. The dissipation effect owing to collisions are neglected 
here. If the effects of dissipation are taken into account $(\omega^2 - \omega_b^2)$ term should be 
replaced with $(\omega^2 - 2i\Gamma \omega - \omega_b^2)$, where $\Gamma \ll \omega$ is the frequency of collision. 

For this work, we assume the photon is traveling in the z-direction. For this case and $B \ll B_{\rm crit}$, the dielectric constants for the vacuum can be written as: 
\begin{eqnarray}
\epsilon_{11} & =  & 1 + {\alpha\over 45 \pi B_{\rm crit}^2}\left(-2 B^2 + 7 B_x B_x\right) \nonumber \\
\epsilon_{22} & = & 1+  {\alpha\over 45 \pi B_{\rm crit}^2}\left(-2 B^2  + 7 B_y B_y \right) \nonumber\\
\epsilon_{12} &  = &  \epsilon_{21}   =   {7\alpha\over 45 \pi B_{\rm crit}^2} B_x B_y
\label{eq:vacbir}
\end{eqnarray}
Here $B^2 = B_x ^2 + B_z^2 + B_y^2$. In the foregoing, we assume the magnetic field to lie in the yz-plane. For this 
case, $\epsilon_{12} = \epsilon_{21} =  0 $. In general, e.g., in another coordinate system,  all three components of the magnetic field would be non-zero and therefore $\epsilon_{12} \ne 0$. It is convenient to discuss the case of $B \ll B_{\rm crit}$ to build
insights into the essential physics. The more general case is not analytically amenable but mostly consistent with
these insights. 

For the plasma, the dielectric components are:
\begin{eqnarray}
\epsilon_{11} & =  & \left(1- {\omega_0^2 \over (\omega^2 - \omega_b^2)} \right) \\
\epsilon_{22} & = & \left(1- {\omega_0^2 \over (\omega^2 - \omega_b^2)} \right)  + \left ({\omega_0^2 \over \omega^2 - \omega_b^2} \right) {\omega_{by}^2 \over \omega^2}   \\
\epsilon_{12} &  = &   - i \left ({\omega_0^2 \over (\omega^2 - \omega_b^2)\omega} \right) \omega_{bz}  \\
\epsilon_{21} &  = &   i \left ({\omega_0^2 \over (\omega^2 - \omega_b^2)\omega} \right) \omega_{bz} 
\end{eqnarray}

We need to assess the importance of vacuum and plasma effects for the typical physical settings we assume in the paper:  $n_e \simeq 10^{12}\hbox{--}10^{23} \, \rm cm^{-3}$ and $B = 10^8\hbox{--}5 \times 10^{14} \, \rm G$. The last scattered photon
arrives from a location which corresponds to $\tau \simeq 1$, where $\tau \simeq \int n_e \sigma_T dl$ and the integral is from the observer to the point of last scattering. After the last scattering, the photon might still traverse regions with high electron densities   and high magnetic fields, which would be determined by  the distribution of electrons and the magnetic field.  Normally, the magnetic field is assumed to be dipolar, but the details of the electron distribution are harder to gauge. 
If the electrons are concentrated in the optically thick accretion column close to the neutron surface, then the last scattered photons emerge from close to the edge of the column and travel in a near-vacuum but high magnetic field environment on the 
way to the observer. In this case, the plasma effects are negligible, but the impact of vacuum birefringence would be significant. 
On the other hand, if the low-density electrons are distributed over larger regions,  then the last scattering occurs far away 
from the neutron surface. In this case, both the plasma and vacuum effects can be negligible. For other configurations, 
both these effects could be important. 

As the Thompson scattering cross section, $\sigma_T$, depends strongly on the energy of the photon in this case,  the photons 
of different frequencies last scatter at different locations. For instance, as noted above,  for $B= 10^{12} \, \rm G$, the scattering cross-section at the resonance is nearly seven orders of magnitude larger than the scattering cross-section without magnetic fields. 
So photons close to the resonance energy last scatter at much smaller densities. At such densities, the impact of plasma 
is negligible. 

We generally find that it is more suitable to neglect plasma effects as compared to the vacuum birefringence. The latter 
cannot be neglected for energies close to the resonance for normal pulsars, which we aim to study here.  Therefore, 
we include the impact of vacuum birefringence here but neglect the possible implications of the plasma birefringence.  

\begin{figure}[h]
    \centering
    \includegraphics[width=0.4\linewidth, keepaspectratio]{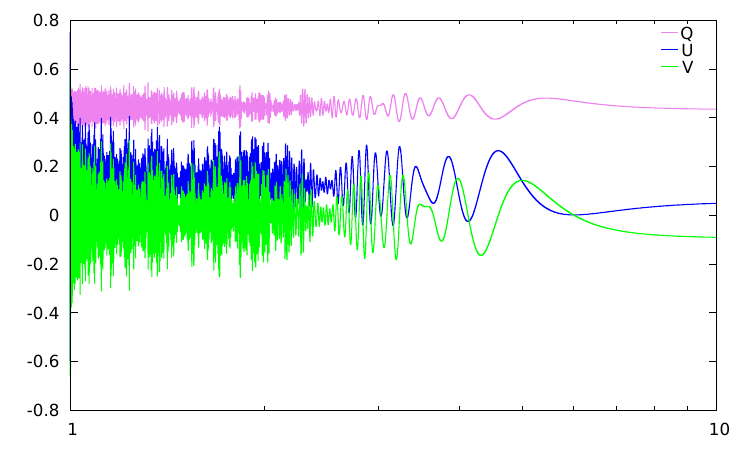}
    \includegraphics[width=0.4\linewidth, keepaspectratio]{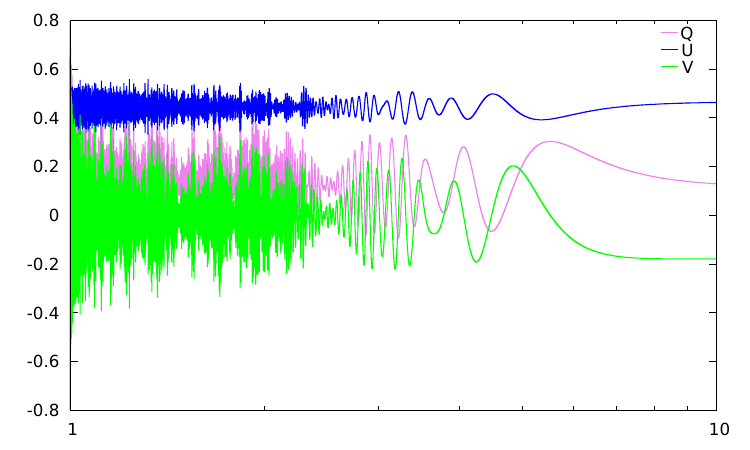}
    \caption{Evolution of Stokes' parameters  ($s_i$) is displayed. The x-axis is in units $z/r$ ($r$ is the position where the photon is last scattered). The panels correspond to the following case: assuming a dipolar magnetic field
    with $B = 10^{12} \, \rm G$ at the surface of the neutron star, the photon is scattered at nearly $7 r_0$, where $r_0$ is the radius of the neutron star.  Initially,  $\Omega_1 =  \Omega_2$ in this case. The initial conditions are: $\{Q, U, V\} =\{0.4, 0.4, 0.8\}$. This gives
    the initial degree of polarization: $\sqrt{s_1^2+s_2^2+s_3^3} = 0.97$. To mimic the situation of mixing of solutions owing to 
    e.g., finite bandwidth, we allow $\Omega_1$ to decrease by nearly two orders of magnitude over 40 steps and average over all the solutions to obtain the final result. In the second panel, the initial conditions are the same, but  $\Omega_2$ is decreased while the other component is held fixed. The difference between the two panels is the reversal of the roles of $Q$ and $U$. Notice that one of the two Stokes' parameters, $Q$ or  $U$, nearly retains its initial condition while the others are driven to zero.  See text for more details.} 
    \label{fig:polrot}
\end{figure}

We follow the formalism developed by \cite{1983JOSA...73.1719K} to study the propagation of polarized light through a vacuum polarized by a strong magnetic field. They showed (see also \citealt{2003ApJ...588..962L,2003PhRvL..91g1101L}): 
\begin{equation}
{d \mathbf{s} \over dz} = \mathbf{\Omega}\times\mathbf{s}
\label{eq:polrot}
\end{equation}
Here $\mathbf{s} = \{Q/I,U/I, V/I\}$, is a three-dimensional vector and $I, Q, U, V$ are Stokes parameters.  
$\mathbf{\Omega}\simeq  k\{(\epsilon_{11}^s - \epsilon_{22}^s), 2\epsilon_{12}^s, 2\epsilon_{12}^a\}$, where superscripts $s$ and $a$ refer
to the symmetric and anti-symmetric parts of the dielectric tensor, and $k = \omega/c$. As the dielectric tensor is symmetric in our
case, $\mathbf{\Omega}_3 = 2\epsilon_{12}^a = 0$. It can be readily verified from  Eq.~(\ref{eq:polrot})   that the evolution of 
Stokes' parameters preserve the
overall degree of polarization: $\sqrt{\mathbf{s}.\mathbf{s}}$. Eq.~(\ref{eq:polrot}) can be expanded: 
\begin{eqnarray}
{ds_1 \over dz} & = &  \Omega_2 s_3 \nonumber \\
{ds_2 \over dz} & = &  -\Omega_1 s_3 \nonumber \\
{ds_3 \over dz} & = &  -\Omega_2 s_1 + \Omega_1 s_2
\label{eq:polrot1}
\end{eqnarray}
We discuss possible solutions of Eqs.~(\ref{eq:polrot1}):
\begin{itemize}
\item[(a)] From the definition of $\mathbf{\Omega}$, it follows that all its components vanish if the magnetic field is aligned
to the direction of propagation of the photon. This is the expected behavior of vacuum birefringence (e.g. \citealt{Gnedin1978}).  From Eqs.~(\ref{eq:vacbir}), only one component of $\mathbf{\Omega}$ is non-vanishing for the specific magnetic field configuration we study the scattering of photons (section~\ref{sec:geocon}).  However, in this subsection, we assume   both  the components of 
$\mathbf{\Omega}$ are non-vanishing with arbitrary magnitudes. 
\item[(b)] We start integrating Eqs.~(\ref{eq:polrot1}) from the position $r$ at which the photon is last scattered.   We expect the initial polarization state of the scattered photon to change substantially if the following condition is met:  
\begin{equation}
k r \alpha \left({B \over B_{\rm crit}} \right )^2 >  1
\end{equation}
Here $r$ is the distance from the neutron star center. If we assume the neutron star to have a dipolar magnetic field, $B =  B_0 r_0^3/r^3$, this condition translates to: 
\begin{equation}
 r  <   \left({7\alpha \mu_0^2 k \over 45 \pi B_{\rm crit}^2} \right)^{1/5} \simeq 4 \times 10^7 \left({\mu_0 \over 10^{30} \, {\rm G cm^{3}}}\right)^{2/5} \left({\nu \over 10 \, \rm keV} \right)^{1/5} \, \rm cm
 \label{eq:rosci}
\end{equation}
Here $\mu_0 = B_0 r_0^3$ is the magnetic moment on the surface of neutron star of radius $r_0 = 10 \, \rm km$.  For a surface magnetic field, $B_0 = 10^{12} \,\rm G$, the impact of vacuum birefringence extends to distances nearly 40 times the radius of the neutron star. 
\item[(c)] For arbitrary $\mathbf{\Omega}$, all three Stokes' parameters display periodic solutions, alternating between positive and negative values. 
\item[(d)] From Eqs.~(\ref{eq:vacbir}) and the definition of $\mathbf{\Omega}$, it follows that  in general one of the components
$\mathbf{\Omega}$  is much smaller than the other. This is expected for most configurations of the magnetic field. For instance, if $B_x \simeq B_y$, $\Omega_1 \ll \Omega_2$, and if $B_{x,y} \gg B_{y,x}$ then
$\Omega_2 \gg \Omega_1$. Therefore, we also consider cases when only one component of $\mathbf{\Omega}$ dominates.  In this case,
the solutions of Eqs.~(\ref{eq:polrot1})   show that one of the Stokes' parameter corresponding to linear polarization nearly retains its initial value while the remaining system of equations display periodic behavior with linear and circular polarization converting into each other. 
\item[(e)] Eqs.~(\ref{eq:polrot1}) determine the evolution of Stokes' parameter for a single frequency. Many other assumptions 
are also implicitly made, e.g., photons arise from the same distance from the neutron star, a homogeneous magnetic field, etc. 
If these additional complications are accounted for, the net outcome would be an average of Stokes' parameters obtained from solutions of  Eqs.~(\ref{eq:polrot1}).  This would  act to drive part of the polarization to zero. For instance,  we need to compute the Stokes' parameter
for a given bandwidth. If the period of the oscillation changes substantially within a bandwidth, then we expect cancellation.  This condition requires: $ (2\pi/c) \delta\nu r  \alpha (B/B_{\rm crit})^2\gg 1$. For X-ray instruments such as IXPE, $(2\pi/c)\delta\nu r  \alpha (B/B_{\rm crit})^2  \gg 1$ is satisfied.  This means we expect strong depolarization within the bandwidth of the instrument (e.g. \citealt{1979ApJ...228L..71C}). Other effects, such as mixing Stokes' parameters
from different locations results in a similar outcome. However, the cancellation only occurs for Stokes' parameters that alternate
between both signs. As discussed above, if one of the components of $\mathbf{\Omega}$ is much smaller than the other, one of the Stokes' parameters corresponding to linear polarization nearly retains its value. This component doesn't suffer cancellation (for detailed
discussion on this and related points see e.g. \citealt{1979ApJ...228L..71C}). This behavior is verified in Figure~\ref{fig:polrot}.
\end{itemize}

Based on the discussion above, we summarize our main findings: 
\begin{itemize}
    \item[(1)] The net impact of vacuum birefringence is to reduce the overall degree of polarization. In particular, it acts to destroy
    the circular polarization and part of the linear polarization. 
    \item[(2)] From Eq.~(\ref{eq:rosci}) we discern the conditions when vacuum birefringence would play an important role. If photons last scattered well above the surface of neutron stars, the impact of birefringence could be neglected. Given our current
    understanding of both accreting and non-accreting neutron stars, it is difficult to determine the point of last scattering. 
    In addition, as already noted above, if the scattered photons emerge close to the direction of the magnetic field, the impact of birefringence could be negligible. 
    \item[(3)] In sum, we expect one of the following outcomes for scattered photons, depending on the impact of vacuum birefringence. 
    The initial degree of polarization would depend on multiple factors. For $B = 10^{12} \, \rm G$,
    close to the resonance frequency, the scattered photons would be strongly circularly polarized  (Figures~\ref{fig:Stokes_unpolarized_incoming},~\ref{fig:Stokes_linear_incoming}, and~\ref{fig:Stokes_circular_incoming}). If the outgoing photons are not closely aligned to the direction of the magnetic field, vacuum polarization effects would 
    destroy the circular polarization, leaving a small residual observable linear polarization. On the other hand, for stronger magnetic fields, the scattered photons would be strongly linearly polarized (e.g., as expected on the surface of magnetars). A fraction of this would survive the vacuum polarization effects, depending on the geometry of the physical setting. This also means that we do expect observed photons from magnetars to have a higher degree of linear polarization compared to normal pulsars, in particular close to the resonance frequency for normal pulsars.
     On the other hand, if the impact of 
    vacuum polarization is negligible, we might see the polarization levels as seen in  Figures~\ref{fig:Stokes_unpolarized_incoming},~\ref{fig:Stokes_linear_incoming}, and~\ref{fig:Stokes_circular_incoming}.  This could become a meaningful diagnostic of the strength and direction of the magnetic field and free particle distribution close to the surface of the neutron stars. A  very distinctive signature in this case could be the degree of polarization as a function of frequency. 
    \item[(4)] {\it Comparison with data}: IXPE data of several pulsars and magnetars is currently available in the frequency range 2--8~keV.  By breaking this spectral range into energy bins, one can also study the spectrum of the polarized emission (for detailed reviews of IXPE observations of  normal pulsars and magnetars see \citealt{Poutanen2024,Taverna2024}). The phase-averaged data doesn't present a simple picture. For instance, the polarization degree for AXP 4U 0142+61,  the brightest among the magnetar candidates, decreases from 15~\% at 2-3~keV to nearly zero at 4--5~keV before increasing to over 35~\% at 6--8~keV.
    Generally, the magnetar emission is seen to be more strongly polarized, with polarization degree increasing from 15--20~\%
    at 2--4~keV to 80~\% in the higher energy bands. The degree of polarization is much lower in pulsars (around 10~\%) with 
    no significant variation with energy, except X~Persei, which showed strong variation with the polarization degree increasing to 30\% in the higher energy band from nearly zero in the lower energy band.  Vela X-1 also shows interesting variation with phase and energy \citep{Forsblom2023}. Given the complexity of the data and uncertainties of theoretical modeling, it is not possible 
    to perform a detailed comparison of our predictions with the data. Our study shows why the pulsars might show lower levels
    of linear polarization as compared to magnetars, which is in conformity with the data. The spectral features accompanying 
    polarized emission shown in Figures~\ref{fig:Stokes_unpolarized_incoming},~\ref{fig:Stokes_linear_incoming}, and~\ref{fig:Stokes_circular_incoming}  cannot be directly compared with the data owing to the possible impact of 
    vacuum polarization, which can only be partially quantified. 
\end{itemize}

\section{Discussion and Conclusion} \label{sec:discussion}
X-ray polarimetry has opened a new window into understanding the physical processes around compact objects such as neutron stars.
In a strongly magnetized neutron star, the polarized emission can arise from Synchrotron radiation from ultra-relativistic particles, which is considered one of the main radiation mechanisms for pulsars. Alternatively, the polarized emission could be owing to the scattering of unpolarized photons from the accretion column, e.g., in a fan beam geometry. This is the favored mechanism
for polarized emission from accreting systems and has been extensively investigated. Our study is a contribution to this literature.
Our main focus in this paper is to investigate the geometry of the incoming light (optically thin or optically thick medium), the dependence of final polarization on the initial polarization of photons,  and the spectral 
properties of the scattered light for a wide range of magnetic field strengths ($10^8 < B < 5 \times 10^{14} \, \rm G$). We 
aim to discern generic properties of the scattered light without complex modeling of the system, e.g., multi-components. 

We summarize our main results and their possible interpretation in light of recent IXPE observations. 
\begin{itemize}
    \item[(a)] For normal pulsars, the observed polarization degree is around $\simeq 5\hbox{--}10$ percent from IXPE data, which can be explained if the scattered photon is roughly aligned  with the magnetic field ($\alpha \lesssim 30$~degrees); both Figures~\ref{fig:Stokes_unpolarized_incoming}
    and~\ref{fig:Stokes_unpolarized_incoming_bothangles} confirm this. The impact of vacuum birefringence is also reduced in this case, but it is not necessarily small.  Eq.~(\ref{eq:rosci}) shows that for $\alpha \simeq 5$~degrees the photon needs to last scatter 5--6 radii above the neutron surface for this effect to be negligible. Even if vacuum polarization plays an important role, we expect residual linear polarization, which might explain the observed degree of polarization. IXPE observations do not show strong dependence on energy, which is harder to explain in the optically thick case but might be possible if the medium were optically thin (Figure~\ref{fig:Stokes_unpolarized_incoming_bothangles}). 
    \item[(b)] For magnetars, as seen in Figures~\ref{fig:Stokes_unpolarized_incoming} and~\ref{fig:Stokes_unpolarized_incoming_bothangles},   our generic prediction is that the observed linear degree of polarization is high, which agrees with the IXPE observations. However, IXPE observations clearly show that the degree of polarization increases with 
    energy, but our predictions do not agree with this observed feature. It is also hard to assess whether such a spectral feature could 
    be created by vacuum birefringence. This is probably an indication that the system is more complicated. For instance, a multi-component system with low energy emission dominated by thermal emission might explain such a feature (e.g., \citealt{Taverna_2026}). It is also conceivable that Synchrotron emission might play a role in such a system. 
\end{itemize}
In our work, we do not consider the time dependence of the observed signal. It is required for comparison with the phase-resolved data. That adds another level of complication, and its modeling requires further assumptions. For example, we could simulate a time-series by expressing $\alpha$ in terms of phase, or a geometric setting similar to the orthogonal rotator model (see e.g., \citealt{Caiazzo2021, Tsygankov2023}). 

Upcoming X-ray polarimeters will expand the energy range probed by IXPE, e.g., eXTP \citep{eXTP}, and POLIX \footnote{https://www.isro.gov.in/XPoSat.html}. These experiments
will be able to probe the predicted features around the resonance frequency.  None of the current or the upcoming experiments
has the capability to detect circular polarization. The relative strength of circular and linear polarization could be 
definitive probe of the scattering paradigm. In addition, as argued above, the circular polarization is unlikely to 
survive if the vacuum birefringence plays an important role. Therefore, the detection of this component 
of polarization might provide meaningful insight into the geometry of scattering and the impact of vacuum birefringence.  However, the measurement of circular polarization in X-ray energy band remains a challenge.

\software{Mathematica (\url{https://www.wolfram.com/mathematica/})
          }




\bibliography{Polarization_in_Neutron_stars}{}

@ARTICLE{Weisskopf2022,
       author = {{Weisskopf}, Martin C. and {Soffitta}, Paolo and {Baldini}, Luca and {Ramsey}, Brian D. and {O'Dell}, Stephen L. and {Romani}, Roger W. and {Matt}, Giorgio and {Deininger}, William D. and {Baumgartner}, Wayne H. and {Bellazzini}, Ronaldo and {Costa}, Enrico and {Kolodziejczak}, Jeffery J. and {Latronico}, Luca and {Marshall}, Herman L. and {Muleri}, Fabio and {Bongiorno}, Stephen D. and {Tennant}, Allyn and {Bucciantini}, Niccolo and {Dovciak}, Michal and {Marin}, Frederic and {Marscher}, Alan and {Poutanen}, Juri and {Slane}, Pat and {Turolla}, Roberto and {Kalinowski}, William and {Di Marco}, Alessandro and {Fabiani}, Sergio and {Minuti}, Massimo and {La Monaca}, Fabio and {Pinchera}, Michele and {Rankin}, John and {Sgro'}, Carmelo and {Trois}, Alessio and {Xie}, Fei and {Alexander}, Cheryl and {Allen}, D. Zachery and {Amici}, Fabrizio and {Andersen}, Jason and {Antonelli}, Angelo and {Antoniak}, Spencer and {Attin{\`a}}, Primo and {Barbanera}, Mattia and {Bachetti}, Matteo and {Baggett}, Randy M. and {Bladt}, Jeff and {Brez}, Alessandro and {Bonino}, Raffaella and {Boree}, Christopher and {Borotto}, Fabio and {Breeding}, Shawn and {Brienza}, Daniele and {Bygott}, H. Kyle and {Caporale}, Ciro and {Cardelli}, Claudia and {Carpentiero}, Rita and {Castellano}, Simone and {Castronuovo}, Marco and {Cavalli}, Luca and {Cavazzuti}, Elisabetta and {Ceccanti}, Marco and {Centrone}, Mauro and {Citraro}, Saverio and {D'Amico}, Fabio and {D'Alba}, Elisa and {Di Gesu}, Laura and {Del Monte}, Ettore and {Dietz}, Kurtis L. and {Di Lalla}, Niccolo' and {Persio}, Giuseppe Di and {Dolan}, David and {Donnarumma}, Immacolata and {Evangelista}, Yuri and {Ferrant}, Kevin and {Ferrazzoli}, Riccardo and {Ferrie}, MacKenzie and {Footdale}, Joseph and {Forsyth}, Brent and {Foster}, Michelle and {Garelick}, Benjamin and {Gunji}, Shuichi and {Gurnee}, Eli and {Head}, Michael and {Hibbard}, Grant and {Johnson}, Samantha and {Kelly}, Erik and {Kilaru}, Kiranmayee and {Lefevre}, Carlo and {Roy}, Shelley Le and {Loffredo}, Pasqualino and {Lorenzi}, Paolo and {Lucchesi}, Leonardo and {Maddox}, Tyler and {Magazzu}, Guido and {Maldera}, Simone and {Manfreda}, Alberto and {Mangraviti}, Elio and {Marengo}, Marco and {Marrocchesi}, Alessandra and {Massaro}, Francesco and {Mauger}, David and {McCracken}, Jeffrey and {McEachen}, Michael and {Mize}, Rondal and {Mereu}, Paolo and {Mitchell}, Scott and {Mitsuishi}, Ikuyuki and {Morbidini}, Alfredo and {Mosti}, Federico and {Nasimi}, Hikmat and {Negri}, Barbara and {Negro}, Michela and {Nguyen}, Toan and {Nitschke}, Isaac and {Nuti}, Alessio and {Onizuka}, Mitch and {Oppedisano}, Chiara and {Orsini}, Leonardo and {Osborne}, Darren and {Pacheco}, Richard and {Paggi}, Alessandro and {Painter}, Will and {Pavelitz}, Steven D. and {Pentz}, Christina and {Piazzolla}, Raffaele and {Perri}, Matteo and {Pesce-Rollins}, Melissa and {Peterson}, Colin and {Pilia}, Maura and {Profeti}, Alessandro and {Puccetti}, Simonetta and {Ranganathan}, Jaganathan and {Ratheesh}, Ajay and {Reedy}, Lee and {Root}, Noah and {Rubini}, Alda and {Ruswick}, Stephanie and {Sanchez}, Javier and {Sarra}, Paolo and {Santoli}, Francesco and {Scalise}, Emanuele and {Sciortino}, Andrea and {Schroeder}, Christopher and {Seek}, Tim and {Sosdian}, Kalie and {Spandre}, Gloria and {Speegle}, Chet O. and {Tamagawa}, Toru and {Tardiola}, Marcello and {Tobia}, Antonino and {Thomas}, Nicholas E. and {Valerie}, Robert and {Vimercati}, Marco and {Walden}, Amy L. and {Weddendorf}, Bruce and {Wedmore}, Jeffrey and {Welch}, David and {Zanetti}, Davide and {Zanetti}, Francesco},
        title = "{The Imaging X-Ray Polarimetry Explorer (IXPE): Pre-Launch}",
      journal = {Journal of Astronomical Telescopes, Instruments, and Systems},
         year = 2022,
        month = apr,
       volume = {8},
       number = {2},
          eid = {026002},
        pages = {026002},
          doi = {10.1117/1.JATIS.8.2.026002},
archivePrefix = {arXiv},
       eprint = {2112.01269},
 primaryClass = {astro-ph.IM},
       adsurl = {https://ui.adsabs.harvard.edu/abs/2022JATIS...8b6002W}
}

@ARTICLE{Doroshenko2022,
       author = {{Doroshenko}, Victor and {Poutanen}, Juri and {Tsygankov}, Sergey S. and {Suleimanov}, Valery F. and {Bachetti}, Matteo and {Caiazzo}, Ilaria and {Costa}, Enrico and {Di Marco}, Alessandro and {Heyl}, Jeremy and {La Monaca}, Fabio and {Muleri}, Fabio and {Mushtukov}, Alexander A. and {Pavlov}, George G. and {Ramsey}, Brian D. and {Rankin}, John and {Santangelo}, Andrea and {Soffitta}, Paolo and {Staubert}, R{\"u}diger and {Weisskopf}, Martin C. and {Zane}, Silvia and {Agudo}, Iv{\'a}n and {Antonelli}, Lucio A. and {Baldini}, Luca and {Baumgartner}, Wayne H. and {Bellazzini}, Ronaldo and {Bianchi}, Stefano and {Bongiorno}, Stephen D. and {Bonino}, Raffaella and {Brez}, Alessandro and {Bucciantini}, Niccol{\`o} and {Capitanio}, Fiamma and {Castellano}, Simone and {Cavazzuti}, Elisabetta and {Ciprini}, Stefano and {De Rosa}, Alessandra and {Del Monte}, Ettore and {Di Gesu}, Laura and {Di Lalla}, Niccol{\`o} and {Donnarumma}, Immacolata and {Dov{\v{c}}iak}, Michal and {Ehlert}, Steven R. and {Enoto}, Teruaki and {Evangelista}, Yuri and {Fabiani}, Sergio and {Ferrazzoli}, Riccardo and {Garcia}, Javier A. and {Gunji}, Shuichi and {Hayashida}, Kiyoshi and {Iwakiri}, Wataru and {Jorstad}, Svetlana G. and {Karas}, Vladimir and {Kitaguchi}, Takao and {Kolodziejczak}, Jeffery J. and {Krawczynski}, Henric and {Latronico}, Luca and {Liodakis}, Ioannis and {Maldera}, Simone and {Manfreda}, Alberto and {Marin}, Fr{\'e}d{\'e}ric and {Marinucci}, Andrea and {Marscher}, Alan P. and {Marshall}, Herman L. and {Matt}, Giorgio and {Mitsuishi}, Ikuyuki and {Mizuno}, Tsunefumi and {Ng}, Chi-Yung and {O'Dell}, Stephen L. and {Omodei}, Nicola and {Oppedisano}, Chiara and {Papitto}, Alessandro and {Peirson}, Abel L. and {Perri}, Matteo and {Pesce-Rollins}, Melissa and {Pilia}, Maura and {Possenti}, Andrea and {Puccetti}, Simonetta and {Ratheesh}, Ajay and {Romani}, Roger W. and {Sgr{\`o}}, Carmelo and {Slane}, Patrick and {Spandre}, Gloria and {Sunyaev}, Rashid A. and {Tamagawa}, Toru and {Tavecchio}, Fabrizio and {Taverna}, Roberto and {Tawara}, Yuzuru and {Tennant}, Allyn F. and {Thomas}, Nicolas E. and {Tombesi}, Francesco and {Trois}, Alessio and {Turolla}, Roberto and {Vink}, Jacco and {Wu}, Kinwah and {Xie}, Fei},
        title = "{Determination of X-ray pulsar geometry with IXPE polarimetry}",
      journal = {Nature Astronomy},
         year = 2022,
        month = dec,
       volume = {6},
        pages = {1433-1443},
          doi = {10.1038/s41550-022-01799-5},
archivePrefix = {arXiv},
       eprint = {2206.07138},
 primaryClass = {astro-ph.HE},
       adsurl = {https://ui.adsabs.harvard.edu/abs/2022NatAs...6.1433D}
}

@ARTICLE{Forsblom2023,
       author = {{Forsblom}, Sofia V. and {Poutanen}, Juri and {Tsygankov}, Sergey S. and {Bachetti}, Matteo and {Di Marco}, Alessandro and {Doroshenko}, Victor and {Heyl}, Jeremy and {La Monaca}, Fabio and {Malacaria}, Christian and {Marshall}, Herman L. and {Muleri}, Fabio and {Mushtukov}, Alexander A. and {Pilia}, Maura and {Rogantini}, Daniele and {Suleimanov}, Valery F. and {Taverna}, Roberto and {Xie}, Fei and {Agudo}, Iv{\'a}n and {Antonelli}, Lucio A. and {Baldini}, Luca and {Baumgartner}, Wayne H. and {Bellazzini}, Ronaldo and {Bianchi}, Stefano and {Bongiorno}, Stephen D. and {Bonino}, Raffaella and {Brez}, Alessandro and {Bucciantini}, Niccol{\`o} and {Capitanio}, Fiamma and {Castellano}, Simone and {Cavazzuti}, Elisabetta and {Chen}, Chien-Ting and {Ciprini}, Stefano and {Costa}, Enrico and {De Rosa}, Alessandra and {Del Monte}, Ettore and {Di Gesu}, Laura and {Di Lalla}, Niccol{\`o} and {Donnarumma}, Immacolata and {Dov{\v{c}}iak}, Michal and {Ehlert}, Steven R. and {Enoto}, Teruaki and {Evangelista}, Yuri and {Fabiani}, Sergio and {Ferrazzoli}, Riccardo and {Garcia}, Javier A. and {Gunji}, Shuichi and {Hayashida}, Kiyoshi and {Iwakiri}, Wataru and {Jorstad}, Svetlana G. and {Kaaret}, Philip and {Karas}, Vladimir and {Kitaguchi}, Takao and {Kolodziejczak}, Jeffery J. and {Krawczynski}, Henric and {Latronico}, Luca and {Liodakis}, Ioannis and {Maldera}, Simone and {Manfreda}, Alberto and {Marin}, Fr{\'e}d{\'e}ric and {Marinucci}, Andrea and {Marscher}, Alan P. and {Matt}, Giorgio and {Mitsuishi}, Ikuyuki and {Mizuno}, Tsunefumi and {Negro}, Michela and {Ng}, Chi-Yung and {O'Dell}, Stephen L. and {Omodei}, Nicola and {Oppedisano}, Chiara and {Papitto}, Alessandro and {Pavlov}, George G. and {Peirson}, Abel L. and {Perri}, Matteo and {Pesce-Rollins}, Melissa and {Petrucci}, Pierre-Olivier and {Possenti}, Andrea and {Puccetti}, Simonetta and {Ramsey}, Brian D. and {Rankin}, John and {Ratheesh}, Ajay and {Roberts}, Oliver J. and {Romani}, Roger W. and {Sgr{\`o}}, Carmelo and {Slane}, Patrick and {Soffitta}, Paolo and {Spandre}, Gloria and {Sunyaev}, Rashid A. and {Swartz}, Douglas A. and {Tamagawa}, Toru and {Tavecchio}, Fabrizio and {Tawara}, Yuzuru and {Tennant}, Allyn F. and {Thomas}, Nicholas E. and {Tombesi}, Francesco and {Trois}, Alessio and {Turolla}, Roberto and {Vink}, Jacco and {Weisskopf}, Martin C. and {Wu}, Kinwah and {Zane}, Silvia and {IXPE Collaboration}},
        title = "{IXPE Observations of the Quintessential Wind-accreting X-Ray Pulsar Vela X-1}",
      journal = {\apjl},
         year = 2023,
        month = apr,
       volume = {947},
       number = {2},
          eid = {L20},
        pages = {L20},
          doi = {10.3847/2041-8213/acc391},
archivePrefix = {arXiv},
       eprint = {2303.01800},
 primaryClass = {astro-ph.HE},
       adsurl = {https://ui.adsabs.harvard.edu/abs/2023ApJ...947L..20F}
}

@ARTICLE{Tsygankov2023,
       author = {{Tsygankov}, Sergey S. and {Doroshenko}, Victor and {Mushtukov}, Alexander A. and {Poutanen}, Juri and {Di Marco}, Alessandro and {Heyl}, Jeremy and {La Monaca}, Fabio and {Forsblom}, Sofia V. and {Malacaria}, Christian and {Marshall}, Herman L. and {Suleimanov}, Valery F. and {Svoboda}, Jiri and {Taverna}, Roberto and {Ursini}, Francesco and {Agudo}, Iv{\'a}n and {Antonelli}, Lucio A. and {Bachetti}, Matteo and {Baldini}, Luca and {Baumgartner}, Wayne H. and {Bellazzini}, Ronaldo and {Bianchi}, Stefano and {Bongiorno}, Stephen D. and {Bonino}, Raffaella and {Brez}, Alessandro and {Bucciantini}, Niccol{\`o} and {Capitanio}, Fiamma and {Castellano}, Simone and {Cavazzuti}, Elisabetta and {Chen}, Chien-Ting and {Ciprini}, Stefano and {Costa}, Enrico and {De Rosa}, Alessandra and {Del Monte}, Ettore and {Di Gesu}, Laura and {Di Lalla}, Niccol{\`o} and {Donnarumma}, Immacolata and {Dov{\v{c}}iak}, Michal and {Ehlert}, Steven R. and {Enoto}, Teruaki and {Evangelista}, Yuri and {Fabiani}, Sergio and {Ferrazzoli}, Riccardo and {Garcia}, Javier A. and {Gunji}, Shuichi and {Hayashida}, Kiyoshi and {Iwakiri}, Wataru and {Jorstad}, Svetlana G. and {Kaaret}, Philip and {Karas}, Vladimir and {Kislat}, Fabian and {Kitaguchi}, Takao and {Kolodziejczak}, Jeffery J. and {Krawczynski}, Henric and {Latronico}, Luca and {Liodakis}, Ioannis and {Maldera}, Simone and {Manfreda}, Alberto and {Marin}, Fr{\'e}d{\'e}ric and {Marinucci}, Andrea and {Marscher}, Alan P. and {Massaro}, Francesco and {Matt}, Giorgio and {Mitsuishi}, Ikuyuki and {Mizuno}, Tsunefumi and {Muleri}, Fabio and {Negro}, Michela and {Ng}, Chi-Yung and {O'Dell}, Stephen L. and {Omodei}, Nicola and {Oppedisano}, Chiara and {Papitto}, Alessandro and {Pavlov}, George G. and {Peirson}, Abel L. and {Perri}, Matteo and {Pesce-Rollins}, Melissa and {Petrucci}, Pierre-Olivier and {Pilia}, Maura and {Possenti}, Andrea and {Puccetti}, Simonetta and {Ramsey}, Brian D. and {Rankin}, John and {Ratheesh}, Ajay and {Roberts}, Oliver J. and {Romani}, Roger W. and {Sgr{\`o}}, Carmelo and {Slane}, Patrick and {Soffitta}, Paolo and {Spandre}, Gloria and {Swartz}, Douglas A. and {Tamagawa}, Toru and {Tavecchio}, Fabrizio and {Tawara}, Yuzuru and {Tennant}, Allyn F. and {Thomas}, Nicholas E. and {Tombesi}, Francesco and {Trois}, Alessio and {Turolla}, Roberto and {Vink}, Jacco and {Weisskopf}, Martin C. and {Wu}, Kinwah and {Xie}, Fei and {Zane}, Silvia},
        title = "{X-ray pulsar GRO J1008‒57 as an orthogonal rotator}",
      journal = {\aap},
         year = 2023,
        month = jul,
       volume = {675},
          eid = {A48},
        pages = {A48},
          doi = {10.1051/0004-6361/202346134},
archivePrefix = {arXiv},
       eprint = {2302.06680},
 primaryClass = {astro-ph.HE},
       adsurl = {https://ui.adsabs.harvard.edu/abs/2023A&A...675A..48T}
}

@ARTICLE{Farinelli2023,
       author = {{Farinelli}, R. and {Fabiani}, S. and {Poutanen}, J. and {Ursini}, F. and {Ferrigno}, C. and {Bianchi}, S. and {Cocchi}, M. and {Capitanio}, F. and {De Rosa}, A. and {Gnarini}, A. and {Kislat}, F. and {Matt}, G. and {Mikusincova}, R. and {Muleri}, F. and {Agudo}, I. and {Antonelli}, L.~A. and {Bachetti}, M. and {Baldini}, L. and {Baumgartner}, W.~H. and {Bellazzini}, R. and {Bongiorno}, S.~D. and {Bonino}, R. and {Brez}, A. and {Bucciantini}, N. and {Castellano}, S. and {Cavazzuti}, E. and {Ciprini}, S. and {Costa}, E. and {Del Monte}, E. and {Di Gesu}, L. and {Di Lalla}, N. and {Di Marco}, A. and {Donnarumma}, I. and {Doroshenko}, V. and {Dov{\v{c}}iak}, M. and {Ehlert}, S.~R. and {Enoto}, T. and {Evangelista}, Y. and {Ferrazzoli}, R. and {Garcia}, J.~A. and {Gunji}, S. and {Hayashida}, K. and {Heyl}, J. and {Iwakiri}, W. and {Jorstad}, S.~G. and {Karas}, V. and {Kitaguchi}, T. and {Kolodziejczak}, J.~J. and {Krawczynski}, H. and {La Monaca}, F. and {Latronico}, L. and {Liodakis}, I. and {Maldera}, S. and {Manfreda}, A. and {Marin}, F. and {Marscher}, A.~P. and {Marshall}, H.~L. and {Mitsuishi}, I. and {Mizuno}, T. and {Ng}, C.-Y. and {O'Dell}, S.~L. and {Omodei}, N. and {Oppedisano}, C. and {Papitto}, A. and {Pavlov}, G.~G. and {Peirson}, A.~L. and {Perri}, M. and {Pesce-Rollins}, M. and {Petrucci}, P.~O. and {Pilia}, M. and {Possenti}, A. and {Puccetti}, S. and {Ramsey}, B.~D. and {Rankin}, J. and {Ratheesh}, A. and {Romani}, R.~W. and {Sgr{\`o}}, C. and {Slane}, P. and {Soffitta}, P. and {Spandre}, G. and {Tamagawa}, T. and {Tavecchio}, F. and {Taverna}, R. and {Tawara}, Y. and {Tennant}, A.~F. and {Thomas}, N.~E. and {Tombesi}, F. and {Trois}, A. and {Tsygankov}, S.~S. and {Turolla}, R. and {Vink}, J. and {Weisskopf}, M.~C. and {Wu}, K. and {Xie}, F. and {Zane}, S.},
        title = "{Accretion geometry of the neutron star low mass X-ray binary Cyg X-2 from X-ray polarization measurements}",
      journal = {\mnras},
         year = 2023,
        month = mar,
       volume = {519},
       number = {3},
        pages = {3681-3690},
          doi = {10.1093/mnras/stac3726},
archivePrefix = {arXiv},
       eprint = {2212.13119},
 primaryClass = {astro-ph.HE},
       adsurl = {https://ui.adsabs.harvard.edu/abs/2023MNRAS.519.3681F}
}

@ARTICLE{Ursini2023,
       author = {{Ursini}, F. and {Farinelli}, R. and {Gnarini}, A. and {Poutanen}, J. and {Bianchi}, S. and {Capitanio}, F. and {Di Marco}, A. and {Fabiani}, S. and {La Monaca}, F. and {Malacaria}, C. and {Matt}, G. and {Miku{\v{s}}incov{\'a}}, R. and {Cocchi}, M. and {Kaaret}, P. and {Kajava}, J.~J.~E. and {Pilia}, M. and {Zhang}, W. and {Agudo}, I. and {Antonelli}, L.~A. and {Bachetti}, M. and {Baldini}, L. and {Baumgartner}, W.~H. and {Bellazzini}, R. and {Bongiorno}, S.~D. and {Bonino}, R. and {Brez}, A. and {Bucciantini}, N. and {Castellano}, S. and {Cavazzuti}, E. and {Chen}, C.-T. and {Ciprini}, S. and {Costa}, E. and {De Rosa}, A. and {Del Monte}, E. and {Di Gesu}, L. and {Di Lalla}, N. and {Donnarumma}, I. and {Doroshenko}, V. and {Dov{\v{c}}iak}, M. and {Ehlert}, S.~R. and {Enoto}, T. and {Evangelista}, Y. and {Ferrazzoli}, R. and {Garcia}, J.~A. and {Gunji}, S. and {Hayashida}, K. and {Heyl}, J. and {Iwakiri}, W. and {Jorstad}, S.~G. and {Karas}, V. and {Kislat}, F. and {Kitaguchi}, T. and {Kolodziejczak}, J.~J. and {Krawczynski}, H. and {Latronico}, L. and {Liodakis}, I. and {Maldera}, S. and {Manfreda}, A. and {Marin}, F. and {Marinucci}, A. and {Marscher}, A.~P. and {Marshall}, H.~L. and {Massaro}, F. and {Mitsuishi}, I. and {Mizuno}, T. and {Muleri}, F. and {Negro}, M. and {Ng}, C.-Y. and {O'Dell}, S.~L. and {Omodei}, N. and {Oppedisano}, C. and {Papitto}, A. and {Pavlov}, G.~G. and {Peirson}, A.~L. and {Perri}, M. and {Pesce-Rollins}, M. and {Petrucci}, P.-O. and {Pilia}, M. and {Possenti}, A. and {Puccetti}, S. and {Ramsey}, B.~D. and {Rankin}, J. and {Ratheesh}, A. and {Roberts}, O.~J. and {Romani}, R.~W. and {Sgr{\`o}}, C. and {Slane}, P. and {Soffitta}, P. and {Spandre}, G. and {Swartz}, D.~A. and {Tamagawa}, T. and {Tavecchio}, F. and {Taverna}, R. and {Tawara}, Y. and {Tennant}, A.~F. and {Thomas}, N.~E. and {Tombesi}, F. and {Trois}, A. and {Tsygankov}, S.~S. and {Turolla}, R. and {Vink}, J. and {Weisskopf}, M.~C. and {Wu}, K. and {Xie}, F. and {Zane}, S.},
        title = "{X-ray polarimetry and spectroscopy of the neutron star low-mass X-ray binary GX 9+9: An in-depth study with IXPE and NuSTAR}",
      journal = {\aap},
         year = 2023,
        month = aug,
       volume = {676},
          eid = {A20},
        pages = {A20},
          doi = {10.1051/0004-6361/202346541},
archivePrefix = {arXiv},
       eprint = {2306.02740},
 primaryClass = {astro-ph.HE},
       adsurl = {https://ui.adsabs.harvard.edu/abs/2023A&A...676A..20U}
}

@ARTICLE{Capitanio2023,
       author = {{Capitanio}, Fiamma and {Fabiani}, Sergio and {Gnarini}, Andrea and {Ursini}, Francesco and {Ferrigno}, Carlo and {Matt}, Giorgio and {Poutanen}, Juri and {Cocchi}, Massimo and {Mikusincova}, Romana and {Farinelli}, Ruben and {Bianchi}, Stefano and {Kajava}, Jari J.~E. and {Muleri}, Fabio and {Sanchez-Fernandez}, Celia and {Soffitta}, Paolo and {Wu}, Kinwah and {Agudo}, Iv{\'a}n and {Antonelli}, Lucio A. and {Bachetti}, Matteo and {Baldini}, Luca and {Baumgartner}, Wayne H. and {Bellazzini}, Ronaldo and {Bongiorno}, Stephen D. and {Bonino}, Raffaella and {Brez}, Alessandro and {Bucciantini}, Niccol{\`o} and {Castellano}, Simone and {Cavazzuti}, Elisabetta and {Ciprini}, Stefano and {Costa}, Enrico and {De Rosa}, Alessandra and {Del Monte}, Ettore and {Di Gesu}, Laura and {Di Lalla}, Niccol{\`o} and {Di Marco}, Alessandro and {Donnarumma}, Immacolata and {Doroshenko}, Victor and {Dov{\v{c}}iak}, Michal and {Ehlert}, Steven R. and {Enoto}, Teruaki and {Evangelista}, Yuri and {Ferrazzoli}, Riccardo and {Garcia}, Javier A. and {Gunji}, Shuichi and {Hayashida}, Kiyoshi and {Heyl}, Jeremy and {Iwakiri}, Wataru and {Jorstad}, Svetlana G. and {Karas}, Vladimir and {Kitaguchi}, Takao and {Kolodziejczak}, Jeffery J. and {Krawczynski}, Henric and {La Monaca}, Fabio and {Latronico}, Luca and {Liodakis}, Ioannis and {Maldera}, Simone and {Manfreda}, Alberto and {Marin}, Fr{\'e}d{\'e}ric and {Marinucci}, Andrea and {Marscher}, Alan P. and {Marshall}, Herman L. and {Mitsuishi}, Ikuyuki and {Mizuno}, Tsunefumi and {Ng}, C.-Y. and {O'Dell}, Stephen L. and {Omodei}, Nicola and {Oppedisano}, Chiara and {Papitto}, Alessandro and {Pavlov}, George G. and {Peirson}, Abel L. and {Perri}, Matteo and {Pesce-Rollins}, Melissa and {Petrucci}, Pierre-Olivier and {Pilia}, Maura and {Possenti}, Andrea and {Puccetti}, Simonetta and {Ramsey}, Brian D. and {Rankin}, John and {Ratheesh}, Ajay and {Romani}, Roger W. and {Sgr{\`o}}, Carmelo and {Slane}, Patrick and {Spandre}, Gloria and {Tamagawa}, Toru and {Tavecchio}, Fabrizio and {Taverna}, Roberto and {Tawara}, Yuzuru and {Tennant}, Allyn F. and {Thomas}, Nicholas E. and {Tombesi}, Francesco and {Trois}, Alessio and {Tsygankov}, Sergey S. and {Turolla}, Roberto and {Vink}, Jacco and {Weisskopf}, Martin C. and {Xie}, Fei and {Zane}, Silvia},
        title = "{Polarization Properties of the Weakly Magnetized Neutron Star X-Ray Binary GS 1826-238 in the High Soft State}",
      journal = {\apj},
         year = 2023,
        month = feb,
       volume = {943},
       number = {2},
          eid = {129},
        pages = {129},
          doi = {10.3847/1538-4357/acae88},
archivePrefix = {arXiv},
       eprint = {2212.12472},
 primaryClass = {astro-ph.HE},
       adsurl = {https://ui.adsabs.harvard.edu/abs/2023ApJ...943..129C}
}

@ARTICLE{Taverna2022,
       author = {{Taverna}, Roberto and {Turolla}, Roberto and {Muleri}, Fabio and {Heyl}, Jeremy and {Zane}, Silvia and {Baldini}, Luca and {Gonz{\'a}lez-Caniulef}, Denis and {Bachetti}, Matteo and {Rankin}, John and {Caiazzo}, Ilaria and {Di Lalla}, Niccol{\`o} and {Doroshenko}, Victor and {Errando}, Manel and {Gau}, Ephraim and {K{\i}rm{\i}z{\i}bayrak}, Demet and {Krawczynski}, Henric and {Negro}, Michela and {Ng}, Mason and {Omodei}, Nicola and {Possenti}, Andrea and {Tamagawa}, Toru and {Uchiyama}, Keisuke and {Weisskopf}, Martin C. and {Agudo}, Ivan and {Antonelli}, Lucio A. and {Baumgartner}, Wayne H. and {Bellazzini}, Ronaldo and {Bianchi}, Stefano and {Bongiorno}, Stephen D. and {Bonino}, Raffaella and {Brez}, Alessandro and {Bucciantini}, Niccol{\`o} and {Capitanio}, Fiamma and {Castellano}, Simone and {Cavazzuti}, Elisabetta and {Ciprini}, Stefano and {Costa}, Enrico and {De Rosa}, Alessandra and {Del Monte}, Ettore and {Di Gesu}, Laura and {Di Marco}, Alessandro and {Donnarumma}, Immacolata and {Dov{\v{c}}iak}, Michal and {Ehlert}, Steven R. and {Enoto}, Teruaki and {Evangelista}, Yuri and {Fabiani}, Sergio and {Ferrazzoli}, Riccardo and {Garcia}, Javier A. and {Gunji}, Shuichi and {Hayashida}, Kiyoshi and {Iwakiri}, Wataru and {Jorstad}, Svetlana G. and {Karas}, Vladimir and {Kitaguchi}, Takao and {Kolodziejczak}, Jeffery J. and {La Monaca}, Fabio and {Latronico}, Luca and {Liodakis}, Ioannis and {Maldera}, Simone and {Manfreda}, Alberto and {Marin}, Fr{\'e}d{\'e}ric and {Marinucci}, Andrea and {Marscher}, Alan P. and {Marshall}, Herman L. and {Matt}, Giorgio and {Mitsuishi}, Ikuyuki and {Mizuno}, Tsunefumi and {Ng}, Stephen C.-Y. and {O{\textquoteright}Dell}, Stephen L. and {Oppedisano}, Chiara and {Papitto}, Alessandro and {Pavlov}, George G. and {Peirson}, Abel L. and {Perri}, Matteo and {Pesce-Rollins}, Melissa and {Pilia}, Maura and {Poutanen}, Juri and {Puccetti}, Simonetta and {Ramsey}, Brian D. and {Ratheesh}, Ajay and {Romani}, Roger W. and {Sgr{\`o}}, Carmelo and {Slane}, Patrick and {Soffitta}, Paolo and {Spandre}, Gloria and {Tavecchio}, Fabrizio and {Tawara}, Yuzuru and {Tennant}, Allyn F. and {Thomas}, Nicholas E. and {Tombesi}, Francesco and {Trois}, Alessio and {Tsygankov}, Sergey S. and {Vink}, Jacco and {Wu}, Kinwah and {Xie}, Fei},
        title = "{Polarized x-rays from a magnetar}",
      journal = {Science},
         year = 2022,
        month = nov,
       volume = {378},
       number = {6620},
        pages = {646-650},
          doi = {10.1126/science.add0080},
archivePrefix = {arXiv},
       eprint = {2205.08898},
 primaryClass = {astro-ph.HE},
       adsurl = {https://ui.adsabs.harvard.edu/abs/2022Sci...378..646T}
}

@ARTICLE{Zane2023,
       author = {{Zane}, Silvia and {Taverna}, Roberto and {Gonz{\'a}lez-Caniulef}, Denis and {Muleri}, Fabio and {Turolla}, Roberto and {Heyl}, Jeremy and {Uchiyama}, Keisuke and {Ng}, Mason and {Tamagawa}, Toru and {Caiazzo}, Ilaria and {Di Lalla}, Niccol{\`o} and {Marshall}, Herman L. and {Bachetti}, Matteo and {La Monaca}, Fabio and {Gau}, Ephraim and {Di Marco}, Alessandro and {Baldini}, Luca and {Negro}, Michela and {Omodei}, Nicola and {Rankin}, John and {Matt}, Giorgio and {Pavlov}, George G. and {Kitaguchi}, Takao and {Krawczynski}, Henric and {Kislat}, Fabian and {Kelly}, Ruth and {Agudo}, Iv{\'a}n and {Antonelli}, Lucio A. and {Baumgartner}, Wayne H. and {Bellazzini}, Ronaldo and {Bianchi}, Stefano and {Bongiorno}, Stephen D. and {Bonino}, Raffaella and {Brez}, Alessandro and {Bucciantini}, Niccol{\`o} and {Capitanio}, Fiamma and {Castellano}, Simone and {Cavazzuti}, Elisabetta and {Chen}, Chieng-Ting and {Ciprini}, Stefano and {Costa}, Enrico and {De Rosa}, Alessandra and {Del Monte}, Ettore and {Di Gesu}, Laura and {Donnarumma}, Immacolata and {Doroshenko}, Victor and {Dov{\v{c}}iak}, Michal and {Ehlert}, Steven R. and {Enoto}, Teruaki and {Evangelista}, Yuri and {Fabiani}, Sergio and {Ferrazzoli}, Riccardo and {Garcia}, Javier A. and {Gunji}, Shuichi and {Hayashida}, Kiyoshi and {Iwakiri}, Wataru and {Jorstad}, Svetlana G. and {Kaaret}, Philip and {Karas}, Vladimir and {Kolodziejczak}, Jeffery J. and {Latronico}, Luca and {Liodakis}, Ioannis and {Maldera}, Simone and {Manfreda}, Alberto and {Marin}, Fr{\'e}d{\'e}ric and {Marinucci}, Andrea and {Marscher}, Alan P. and {Massaro}, Francesco and {Mitsuishi}, Ikuyuki and {Mizuno}, Tsunefumi and {Ng}, C.-Y. and {O'Dell}, Stephen L. and {Oppedisano}, Chiara and {Papitto}, Alessandro and {Peirson}, Abel L. and {Perri}, Matteo and {Pesce-Rollins}, Melissa and {Petrucci}, Pierre-Olivier and {Pilia}, Maura and {Possenti}, Andrea and {Poutanen}, Juri and {Puccetti}, Simonetta and {Ramsey}, Brian D. and {Ratheesh}, Ajay and {Roberts}, Oliver J. and {Romani}, Roger W. and {Sgr{\'o}}, Carmelo and {Slane}, Patrick and {Soffitta}, Paolo and {Spandre}, Gloria and {Swartz}, Douglas A. and {Tavecchio}, Fabrizio and {Tawara}, Yuzuru and {Tennant}, Allyn F. and {Thomas}, Nicholas E. and {Tombesi}, Francesco and {Trois}, Alessio and {Tsygankov}, Sergey S. and {Vink}, Jacco and {Weisskopf}, Martin C. and {Wu}, Kinwah and {Xie}, Fei},
        title = "{A Strong X-Ray Polarization Signal from the Magnetar 1RXS J170849.0-400910}",
      journal = {\apjl},
         year = 2023,
        month = feb,
       volume = {944},
       number = {2},
          eid = {L27},
        pages = {L27},
          doi = {10.3847/2041-8213/acb703},
archivePrefix = {arXiv},
       eprint = {2301.12919},
 primaryClass = {astro-ph.HE},
       adsurl = {https://ui.adsabs.harvard.edu/abs/2023ApJ...944L..27Z}
}

@ARTICLE{Caiazzo2021,
       author = {{Caiazzo}, Ilaria and {Heyl}, Jeremy},
        title = "{Polarization of accreting X-ray pulsars. I. A new model}",
      journal = {\mnras},
         year = 2021,
        month = jan,
       volume = {501},
       number = {1},
        pages = {109-128},
          doi = {10.1093/mnras/staa3428},
archivePrefix = {arXiv},
       eprint = {2009.00631},
 primaryClass = {astro-ph.HE},
       adsurl = {https://ui.adsabs.harvard.edu/abs/2021MNRAS.501..109C}
}

@ARTICLE{Meszaros1988,
       author = {{Meszaros}, P. and {Novick}, R. and {Szentgyorgyi}, A. and {Chanan}, G.~A. and {Weisskopf}, M.~C.},
        title = "{Astrophysical Implications and Observational Prospects of X-Ray Polarimetry}",
      journal = {\apj},
         year = 1988,
        month = jan,
       volume = {324},
        pages = {1056},
          doi = {10.1086/165962},
       adsurl = {https://ui.adsabs.harvard.edu/abs/1988ApJ...324.1056M}
}

@ARTICLE{Poutanen2024,
       author = {{Poutanen}, Juri and {Tsygankov}, Sergey S. and {Forsblom}, Sofia V.},
        title = "{X-ray Polarimetry of X-ray Pulsars}",
      journal = {Galaxies},
         year = 2024,
        month = aug,
       volume = {12},
       number = {4},
          eid = {46},
        pages = {46},
          doi = {10.3390/galaxies12040046},
archivePrefix = {arXiv},
       eprint = {2408.04431},
 primaryClass = {astro-ph.HE},
       adsurl = {https://ui.adsabs.harvard.edu/abs/2024Galax..12...46P}
}

@ARTICLE{Taverna2024,
       author = {{Taverna}, Roberto and {Turolla}, Roberto},
        title = "{X-ray Polarization from Magnetar Sources}",
      journal = {Galaxies},
         year = 2024,
        month = feb,
       volume = {12},
       number = {1},
          eid = {6},
        pages = {6},
          doi = {10.3390/galaxies12010006},
archivePrefix = {arXiv},
       eprint = {2402.05622},
 primaryClass = {astro-ph.HE},
       adsurl = {https://ui.adsabs.harvard.edu/abs/2024Galax..12....6T}
}

@ARTICLE{Gnedin1978,
       author = {{Gnedin}, Yu. N. and {Pavlov}, G.~G. and {Shibanov}, Yu. A.},
        title = "{The effect of vacuum birefringence in a magnetic field on the polarization and beaming of X-ray pulsars}",
      journal = {Soviet Astronomy Letters},
         year = 1978,
        month = jan,
       volume = {4},
        pages = {117-119},
       adsurl = {https://ui.adsabs.harvard.edu/abs/1978SvAL....4..117G}
}

@ARTICLE{Heyl2018,
       author = {{Heyl}, Jeremy and {Caiazzo}, Ilaria},
        title = "{Strongly Magnetized Sources: QED and X-ray Polarization}",
      journal = {Galaxies},
         year = 2018,
        month = jul,
       volume = {6},
       number = {3},
          eid = {76},
        pages = {76},
          doi = {10.3390/galaxies6030076},
archivePrefix = {arXiv},
       eprint = {1802.00358},
 primaryClass = {astro-ph.HE},
       adsurl = {https://ui.adsabs.harvard.edu/abs/2018Galax...6...76H}
}

@ARTICLE{Sokolova-Lapa2023,
       author = {{Sokolova-Lapa}, E. and {Stierhof}, J. and {Dauser}, T. and {Wilms}, J.},
        title = "{Vacuum polarization alters the spectra of accreting X-ray pulsars}",
      journal = {\aap},
         year = 2023,
        month = jun,
       volume = {674},
          eid = {L2},
        pages = {L2},
          doi = {10.1051/0004-6361/202346265},
archivePrefix = {arXiv},
       eprint = {2305.00475},
 primaryClass = {astro-ph.HE},
       adsurl = {https://ui.adsabs.harvard.edu/abs/2023A&A...674L...2S}
}

@ARTICLE{Becker2012,
       author = {{Becker}, P.~A. and {Klochkov}, D. and {Sch{\"o}nherr}, G. and {Nishimura}, O. and {Ferrigno}, C. and {Caballero}, I. and {Kretschmar}, P. and {Wolff}, M.~T. and {Wilms}, J. and {Staubert}, R.},
        title = "{Spectral formation in accreting X-ray pulsars: bimodal variation of the cyclotron energy with luminosity}",
      journal = {\aap},
         year = 2012,
        month = aug,
       volume = {544},
          eid = {A123},
        pages = {A123},
          doi = {10.1051/0004-6361/201219065},
archivePrefix = {arXiv},
       eprint = {1205.5316},
 primaryClass = {astro-ph.HE},
       adsurl = {https://ui.adsabs.harvard.edu/abs/2012A&A...544A.123B}
}

@ARTICLE{1983JOSA...73.1719K,
       author = {{Kubo}, Hayao and {Nagata}, Ryo},
        title = "{Vector representation of behavior of polarized light in a weakly inhomogeneous medium with birefringence and dichroism}",
      journal = {Journal of the Optical Society of America (1917-1983)},
         year = 1983,
        month = dec,
       volume = {73},
       number = {12},
        pages = {1719},
          doi = {10.1364/JOSA.73.001719},
       adsurl = {https://ui.adsabs.harvard.edu/abs/1983JOSA...73.1719K}
}

@ARTICLE{1997JPhA...30.6485H,
       author = {{Heyl}, Jeremy S. and {Hernquist}, Lars},
        title = "{Birefringence and dichroism of the QED vacuum}",
      journal = {Journal of Physics A Mathematical General},
         year = 1997,
        month = sep,
       volume = {30},
       number = {18},
        pages = {6485-6492},
          doi = {10.1088/0305-4470/30/18/022},
archivePrefix = {arXiv},
       eprint = {hep-ph/9705367},
 primaryClass = {astro-ph},
       adsurl = {https://ui.adsabs.harvard.edu/abs/1997JPhA...30.6485H}
}

@ARTICLE{2003ApJ...588..962L,
       author = {{Lai}, Dong and {Ho}, Wynn C.~G.},
        title = "{Transfer of Polarized Radiation in Strongly Magnetized Plasmas and Thermal Emission from Magnetars: Effect of Vacuum Polarization}",
      journal = {\apj},
         year = 2003,
        month = may,
       volume = {588},
       number = {2},
        pages = {962-974},
          doi = {10.1086/374334},
archivePrefix = {arXiv},
       eprint = {astro-ph/0211315},
 primaryClass = {astro-ph},
       adsurl = {https://ui.adsabs.harvard.edu/abs/2003ApJ...588..962L}
}

@ARTICLE{2003PhRvL..91g1101L,
       author = {{Lai}, Dong and {Ho}, Wynn C.},
        title = "{Polarized X-Ray Emission from Magnetized Neutron Stars: Signature of Strong-Field Vacuum Polarization}",
      journal = {\prl},
         year = 2003,
        month = aug,
       volume = {91},
       number = {7},
          eid = {071101},
        pages = {071101},
          doi = {10.1103/PhysRevLett.91.071101},
archivePrefix = {arXiv},
       eprint = {astro-ph/0303596},
 primaryClass = {astro-ph},
       adsurl = {https://ui.adsabs.harvard.edu/abs/2003PhRvL..91g1101L}
}

@BOOK{1970pewp.book.....G,
       author = {{Ginzburg}, V.~L.},
        title = "{The propagation of electromagnetic waves in plasmas}",
         year = 1970,
       adsurl = {https://ui.adsabs.harvard.edu/abs/1970pewp.book.....G}
}

@ARTICLE{1978PhRvD..18.1053D,
       author = {{Daugherty}, J.~K. and {Ventura}, J.},
        title = "{Absorption of radiation by electrons in intense magnetic fields}",
      journal = {\prd},
         year = 1978,
        month = aug,
       volume = {18},
       number = {4},
        pages = {1053-1067},
          doi = {10.1103/PhysRevD.18.1053},
       adsurl = {https://ui.adsabs.harvard.edu/abs/1978PhRvD..18.1053D}
}

@BOOK{1998clel.book.....J,
       author = {{Jackson}, John David},
        title = "{Classical Electrodynamics, 3rd Edition}",
         year = 1998,
       adsurl = {https://ui.adsabs.harvard.edu/abs/1998clel.book.....J}
}

@BOOK{1986rpa..book.....R,
       author = {{Rybicki}, George B. and {Lightman}, Alan P.},
        title = "{Radiative Processes in Astrophysics}",
         year = 1986,
       adsurl = {https://ui.adsabs.harvard.edu/abs/1986rpa..book.....R}
}

@ARTICLE{1971PhRvD...3.2303C,
       author = {{Canuto}, V. and {Lodenquai}, J. and {Ruderman}, M.},
        title = "{Thomson Scattering in a Strong Magnetic Field}",
      journal = {\prd},
         year = 1971,
        month = may,
       volume = {3},
       number = {10},
        pages = {2303-2308},
          doi = {10.1103/PhysRevD.3.2303},
       adsurl = {https://ui.adsabs.harvard.edu/abs/1971PhRvD...3.2303C}
}

@ARTICLE{1986Ap&SS.121..333C,
       author = {{Chou}, C.~K.},
        title = "{Stokes Parameters for Thomson Scattering in a Strong Magnetic Field}",
      journal = {\apss},
         year = 1986,
        month = apr,
       volume = {121},
       number = {2},
        pages = {333-344},
          doi = {10.1007/BF00653705},
       adsurl = {https://ui.adsabs.harvard.edu/abs/1986Ap&SS.121..333C}
}

@BOOK{1971ctf..book.....L,
       author = {{Landau}, Lev Davidovich and {Lifshitz}, E.~M.},
        title = "{The classical theory of fields}",
         year = 1971,
       adsurl = {https://ui.adsabs.harvard.edu/abs/1971ctf..book.....L}
}

@ARTICLE{1979ApJ...228L..71C,
       author = {{Chanan}, G.~A. and {Novick}, R. and {Silver}, E.~H.},
        title = "{Further comments on the effects of vacuum birefringence on the polarization of X-rays emitted from magnetic neutron stars.}",
      journal = {\apjl},
         year = 1979,
        month = mar,
       volume = {228},
        pages = {L71-L74},
          doi = {10.1086/182906},
       adsurl = {https://ui.adsabs.harvard.edu/abs/1979ApJ...228L..71C}
}

@article{Taverna_2026,
doi = {10.3847/1538-4357/ae5c9d},
url = {https://doi.org/10.3847/1538-4357/ae5c9d},
year = {2026},
month = {apr},
publisher = {The American Astronomical Society},
volume = {1002},
number = {1},
pages = {102},
author = {Taverna, Roberto and Turolla, Roberto and Marra, Lorenzo and Kelly, Ruth M.E. and Borghese, Alice and Israel, Gian Luca and Mereghetti, Sandro and Possenti, Andrea and Zane, Silvia and Rigoselli, Michela},
title = {The Long Quest for Vacuum Birefringence in Magnetars: 1E 1547.0–5408 and the Elusive Smoking Gun},
journal = {The Astrophysical Journal}
}

@ARTICLE{eXTP,
       author = {{Zhang}, Shuang-Nan and {Santangelo}, Andrea and {Xu}, Yupeng and {Feng}, Hua and {Lu}, Fangjun and {Chen}, Yong and {Ge}, Mingyu and {Nandra}, Kirpal and {Wu}, Xin and {Feroci}, Marco and {Hernanz}, Margarita and {Liu}, Congzhan and {He}, Huilin and {Wang}, Yusa and {Jiang}, Weichun and {Cui}, Weiwei and {Yang}, Yanji and {Wang}, Juan and {Li}, Wei and {Li}, Hong and {Du}, Yuanyuan and {Liu}, Xiaohua and {Meng}, Bin and {Wen}, Xiangyang and {Zhang}, Aimei and {Ma}, Jia and {Li}, Maoshun and {Li}, Gang and {Qi}, Liqiang and {Sun}, Jianchao and {Luo}, Tao and {Liu}, Hongwei and {Liu}, Xiaojing and {Zhang}, Fan and {Luo}, Laidan and {Zhu}, Yuxuan and {Zhao}, Zijian and {Sun}, Liang and {Yang}, Xiongtao and {Wu}, Qiong and {Jiang}, Jiechen and {Shi}, Haoli and {Liu}, Jiangtao and {Xu}, Yanbing and {Yang}, Sheng and {Zhang}, Laiyu and {Han}, Dawei and {Gao}, Na and {Huo}, Jia and {Zhang}, Ziliang and {Wang}, Hao and {Zhao}, Xiaofan and {Wang}, Shuo and {Li}, Zhenjie and {Bao}, Ziyu and {Liu}, Yaoguang and {Wang}, Ke and {Wang}, Na and {Wang}, Bo and {Wang}, Langping and {Wang}, Dianlong and {Ding}, Fei and {Sheng}, Lizhi and {Qiang}, Pengfei and {Yan}, Yongqing and {Liu}, Yongan and {Wu}, Zhenyu and {Liu}, Yichen and {Chen}, Hao and {Zhang}, Yacong and {Liu}, Hongbang and {Altmann}, Alexander and {Bechteler}, Thomas and {Burwitz}, Vadim and {Fiorini}, Carlo and {Friedrich}, Peter and {Meidinger}, Norbert and {Strecker}, Rafael and {Baldini}, Luca and {Bellazzini}, Ronaldo and {Bonino}, Raffaella and {Frass{\`a}}, Andrea and {Latronico}, Luca and {Maldera}, Simone and {Manfreda}, Alberto and {Minuti}, Massimo and {Pesce-Rollins}, Melissa and {Sgr{\`o}}, Carmelo and {Tugliani}, Stefano and {Pareschi}, Giovanni and {Basso}, Stefano and {Sironi}, Giorgia and {Spiga}, Daniele and {Tagliaferri}, Gianpiero and {Tykhonov}, Andrii and {Paltani}, St{\`e}phane and {Bozzo}, Enrico and {Tenzer}, Christoph and {Bayer}, J{\"o}rg and {Tuo}, Youli and {Liu}, Honghui and {Zhang}, Yonghe and {Cai}, Zhiming and {Liu}, Huaqiu and {Chen}, Wen and {Wang}, Chunhong and {He}, Tao and {Chen}, Yehai and {Qiu}, Chengbo and {Zhang}, Ye and {Feng}, Jianchao and {Zhu}, Xiaofei and {Zhou}, Heng and {Zheng}, Shijie and {Song}, Liming and {Wang}, Jinzhou and {Jia}, Shumei and {Jiang}, Zewen and {Li}, Xiaobo and {Zhao}, Haisheng and {Guan}, Ju and {Zhang}, Juan and {Li}, Chengkui and {Huang}, Yue and {Liao}, Jinyuan and {You}, Yuan and {Zhang}, Hongmei and {Wang}, Wenshuai and {Wang}, Shuang and {Ou}, Ge and {Hu}, Hao and {Shi}, Jingyan and {Cui}, Tao and {Jiang}, Xiaowei and {Cheng}, Yaodong and {Li}, Haibo and {Xu}, Yanjun and {Zane}, Silvia and {Bambi}, Cosimo and {Bu}, Qingcui and {Dall'Osso}, Simone and {Rosa}, Alessandra De and {Gou}, Lijun and {Guillot}, Sebastien and {Ji}, Long and {Li}, Ang and {Mao}, Jirong and {Patruno}, Alessandro and {Stratta}, Giulia and {Taverna}, Roberto and {Tsygankov}, Sergey and {Uttley}, Phil and {Watts}, Anna L. and {Wu}, Xuefeng and {Xu}, Renxin and {Yi}, Shuxu and {Zhang}, Guobao and {Zhang}, Liang and {Zhao}, Wen and {Zhou}, Ping},
        title = "{The enhanced X-ray Timing and Polarimetry mission{\textemdash}eXTP for launch in 2030}",
      journal = {Science China Physics, Mechanics, and Astronomy},
         year = 2025,
        month = sep,
       volume = {68},
       number = {11},
          eid = {119502},
        pages = {119502},
          doi = {10.1007/s11433-025-2786-6},
archivePrefix = {arXiv},
       eprint = {2506.08101},
 primaryClass = {astro-ph.HE},
       adsurl = {https://ui.adsabs.harvard.edu/abs/2025SCPMA..6819502Z}
}

@BOOK{1979rpa..book.....R,
       author = {{Rybicki}, George B. and {Lightman}, Alan P.},
        title = "{Radiative processes in astrophysics}",
         year = 1979,
       adsurl = {https://ui.adsabs.harvard.edu/abs/1979rpa..book.....R}
}

@ARTICLE{Goldreich,
       author = {{Goldreich}, Peter and {Julian}, William H.},
        title = "{Pulsar Electrodynamics}",
      journal = {\apj},
         year = 1969,
        month = aug,
       volume = {157},
        pages = {869},
          doi = {10.1086/150119},
       adsurl = {https://ui.adsabs.harvard.edu/abs/1969ApJ...157..869G}
}

@ARTICLE{Harding1991,
       author = {{Harding}, Alice K. and {Daugherty}, Joseph K.},
        title = "{Cyclotron Resonant Scattering and Absorption}",
      journal = {\apj},
         year = 1991,
        month = jun,
       volume = {374},
        pages = {687},
          doi = {10.1086/170153},
       adsurl = {https://ui.adsabs.harvard.edu/abs/1991ApJ...374..687H}
}
\bibliographystyle{aasjournalv7}



\end{document}